\documentclass[pre,aps,twocolumn,showpacs,floatfix,10pt,superscriptaddress]{revtex4-2}
\usepackage{graphicx}
\usepackage{color}
\usepackage{lingmacros}
\usepackage{tree-dvips}
\usepackage{amsmath}
\usepackage{amssymb}
\usepackage{float}
\usepackage{graphicx}
\usepackage{latexsym}
\usepackage{setspace}
\usepackage{wrapfig}
\usepackage[bottom]{footmisc}
\usepackage{bm}
\newcommand{\be}{\begin{equation}}
\newcommand{\ee}{\end{equation}}
\newcommand{\bea}{\begin{eqnarray}}
\newcommand{\eea}{\end{eqnarray}}

\begin{document}
\title{Rejection-free cluster Wang-Landau algorithm for hard-core lattice gases}
\author{Asweel Ahmed A. Jaleel}
\email{asweel@gmail.com}
\affiliation{The Institute of Mathematical Sciences, C.I.T. Campus,
Taramani, Chennai 600113, India}
\affiliation{Homi Bhabha National Institute, Training School Complex, Anushakti Nagar, Mumbai 400094, India}
\author{Jetin E. Thomas}
\email{jetinthomas@imsc.res.in} 
\affiliation{The Institute of Mathematical Sciences, C.I.T. Campus,
Taramani, Chennai 600113, India}
\affiliation{Homi Bhabha National Institute, Training School Complex, Anushakti Nagar, Mumbai 400094, India}
\author{Dipanjan Mandal}
\email{dipanjan.mandal@warwick.ac.uk}
\affiliation{Department of Physics, University of Warwick, Coventry CV4 7AL, United Kingdom}
\author{Sumedha}
\email{sumedha@niser.ac.in}
\affiliation{School of Physical Sciences, National Institute of Science Education and Research, Bhubaneswar, P.O. Jatni, Khurda, Odisha 752050, India}
\affiliation{Homi Bhabha National Institute, Training School Complex, Anushakti Nagar, Mumbai 400094, India}

\author{R. Rajesh}
\email{rrajesh@imsc.res.in}
\affiliation{The Institute of Mathematical Sciences, C.I.T. Campus,
	Taramani, Chennai 600113, India}
\affiliation{Homi Bhabha National Institute, Training School Complex, Anushakti Nagar, Mumbai 400094, India}

%\date{October 25, 2021}
\date{\today}

\begin{abstract}
We introduce a  rejection-free, flat histogram, cluster algorithm to determine the density of states of hard-core lattice gases. We show that the algorithm is able to efficiently sample low entropy states that are usually difficult to access, even when the excluded volume per particle is large. The algorithm is based on simultaneously evaporating all the particles in a strip and reoccupying these sites with a new appropriately chosen configuration. We implement the algorithm for the particular case of the hard-core lattice gas in which the first $k$ next-nearest neighbors of a particle are excluded from being occupied. It is shown that the algorithm is able to reproduce the known results for $k=1,2,3$  both on the square and cubic lattices. We also show that, in comparison, the corresponding flat histogram algorithms with either local moves or unbiased cluster moves are less accurate and do not converge as the system size increases.
\end{abstract}

%\pacs{64.60.Cn,64.70.mf,05.50.+q}

\maketitle

\section{\label{sec:intro}Introduction}

Lattice gas models of particles that interact only through
excluded volume interactions, also known as hard-core lattice gases (HCLGs), are among the simplest systems that undergo phase transitions~\cite{runnels1972phase}. Since the interaction energy is either infinity or zero, depending on whether particles overlap or not, temperature plays no role in causing phase transitions. Any phase transition induced by changing density is driven by a gain in entropy and thus HCLGs are the minimal models for studying entropy-driven transitions. HCLGs are also closely related to the freezing transition~\cite{1957-aw-jcp-phase,1962-aw-pr-phase}, self assembly~\cite{cuetos-2007-phase,rafael-2020-self}, adsorption on surfaces~\cite{1985-prb-twpbe-two,1985-bkuoabe-prl-phase}, directed and undirected lattice animals~\cite{1982-d-prl-equivalence,1983-d-prl-exact,2003-bi-jsp-dimensional}, and the Yang-Lee edge singularity~\cite{1981-ps-prl-critical}. Systems of many differently shaped particles have been studied.  Examples include rods~\cite{1956-f-prs-phase,2007-gd-epl-on,2013-krds-pre-nematic,2017-gkao-pre-isotropic,2017-vdr-jsm-different},  tetrominoes~\cite{2002-mhs-jcp-simple,2009-bsg-langmuir-structure}, triangles~\cite{1999-vn-prl-triangular}, 
 Y-shaped particles~\cite{szabelski2013selfassembly,2015-rthrg-tsf-impact,2018-pre-mnr-phase}, hexagons~\cite{1980-b-jpa-hard}, cubes~\cite{vigneshwar2019phase}, rectangles~\cite{2014-kr-pre-phase,2015-kr-epjb-phase}, discretized spheres~\cite{2007-fal-jcp-monte,2014-nr-pre-multiple,PhysRevE.101.062138,akimenko2019tensor}, etc.

Despite its wide applicability and long history dating back to the 1950s~\cite{1956-f-prs-phase,1958-d-nc-theoretical,1960-b-pps-lattice,1967-bn-jcp-phase,1966-bn-prl-phase,1961-k-physica-statistics}, basic issues like predicting the phases and their order of appearance,   given the shape of the particles, are not satisfactorily resolved. Exact solutions are limited to the case of hard hexagons~\cite{1980-b-jpa-hard}.   Given the analytical intractability, the main tool in studying these systems is Monte Carlo simulations. Conventional Monte Carlo simulations that use local evaporation, deposition, and diffusion moves work well for only low densities. At higher densities or when the excluded volume becomes larger, it becomes difficult to equilibrate the system because the system gets trapped in long-lived metastable states. This difficulty has been overcome by a recently introduced cluster algorithm~\cite{2012-krds-aipcp-monte,2013-krds-pre-nematic,2014-nr-pre-multiple,2015-rdd-prl-columnar} which has been efficient in equilibration even at full packing~\cite{2015-rdd-prl-columnar}, resulting in obtaining the accurate phase diagram of different systems, both in two and three dimensions. We will refer to this algorithm as the strip cluster update algorithm. The basic move in this algorithm is the evaporation of all the particles in a  randomly chosen strip of lattice sites, and reoccupying the strip with a new configuration.  The probabilities of the allowed new configurations are determined from transfer matrix calculations. In this paper, we modify this grand canonical strip update algorithm to obtain the density of states of hard-core lattice gases.

An important development in Monte Carlo simulations is the use of flat histogram algorithms to obtain directly the density of states.  Such methods have a great advantage over conventional Monte Carlo simulations where, for each set of couplings (like temperature, fugacity, field strengths, etc.), the simulation has to be separately done. In flat histogram methods, the density of states can be used to generate data for any value of the coupling.  Some of the early methods used to generate density of states are  multicanonical ensemble method~\cite{berg1992multicanonical}, entropic sampling~\cite{lee1993new}, broad histogram method~\cite{herrmann1996broad} and flat histogram method~\cite{wang2000monte}. The Wang-Landau (WL) algorithm~\cite{2001-wl-prl-efficient,2001-wl-pre-determining} is a very popular flat histogram method in which the density of states evolves continuously during the simulations, resulting in fast convergence of the density of states to its final values. The WL algorithm also overcomes critical slowing down and long relaxation times~\cite{zhou2005understanding}. A review of the algorithm and its applications can be found in Ref.~\cite{singh2012density}. Several variants of the WL method, such as adaptive windows~\cite{cunha2011critical}, 1/t algorithm~\cite{belardinelli2007fast, belardinelli2007wang}, tomographic sampling~\cite{2011-dc-pre-complete}, etc., have also been proposed.

In this paper, we propose a rejection-free strip cluster Wang-Landau (SCWL) update algorithm, combining both the strip cluster update algorithm and the WL algorithm, to determine the density of states of HCLGs. In the evaporation-deposition part of the algorithm, all particles in a strip are removed and reoccupied with a new configuration. Even though we have updated only a single row (strip of width one) at a time for all the models considered in this paper,  in general the choice of width of the strip depends on the model being considered. For example, in mixtures of $2 \times 2$ squares and dimers at full packing, the minimal width is a strip of width two~\cite{2015-rdd-prl-columnar}. The new configurations are chosen in proportion to their weights, which in turn are determined by the current density of states, making the algorithm rejection free. By comparing the performance of SCWL with corresponding algorithms with either single site updates or where new configurations are chosen independent of their weight, we show that both the cluster move as well as the rejection-free choice of new configurations are important to obtain an accurate estimate of the density of states for  HCLGs. As a concrete example, we apply the algorithm to the $k$-NN exclusion model in which a particle excludes all sites up to the $k$th nearest neighbors from being occupied. 
We show that we are able to reproduce the known results for the critical behavior of this model for $k=1, 2, 3$, both on square and cubic lattices. 
For the first order transitions, we show that the nonconvexity of the measured entropy can be utilized to obtain accurate estimates of both the critical chemical potential as well as the coexistence densities.  In the case of the $3$-NN model in two dimensions and the $2$-NN model in three dimensions, we obtain improved estimates for critical chemical potential, coexistence  densities, and critical pressure. The improved estimates of critical chemical potential are $3.6766(5)$ for the $3$-NN model in two dimensions and $0.5326(4)$ for the $2$-NN model in three dimensions. The coexistence densities range from $0.8055(3)$ to $0.9570(3)$ for the $3$-NN model in two dimensions and $0.4136(1)$ to $0.5197(2)$ for the $2$-NN model in three dimensions. The critical pressure is $0.74147(6)$ for the $3$-NN model in two dimensions and $0.2542(1)$ for the $2$-NN model in three dimensions.

The remainder of the paper is organized as follows. In Sec.~\ref{sec:model}, we define the $k$-NN model.  In Sec.~\ref{sec:model b},  we describe the SCWL algorithm as well as variants of the algorithm with either local moves or unbiased evaporation-deposition moves. Section~\ref{sec:comparisons} contains a detailed comparison of the performance of the different variants of the flat histogram algorithms in obtaining the density of states for the 1-NN and 2-NN models in two dimensions. In Sec.~\ref{sec:applications}, the algorithm is applied to the $k$-NN model in two and three dimensions for $k=1,2,3$. The critical behavior of each of these models is obtained.  Finally, we summarize and discuss the relevance of our results in  Sec.~\ref{sec:summary}.

\section{\label{sec:model}$k$-NN hard-core lattice gas}

\begin{figure}
	\includegraphics[width=0.8\columnwidth]{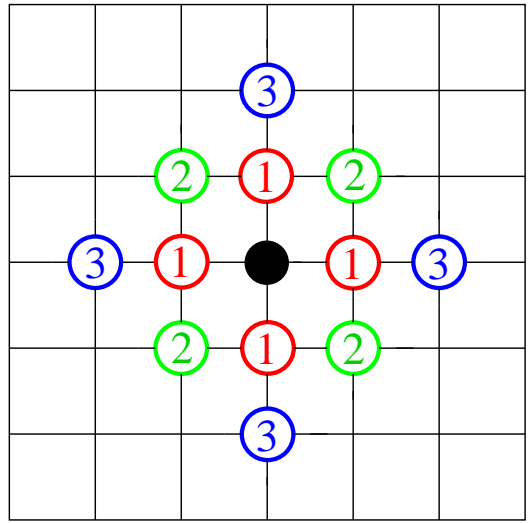}
	\caption{(Color online) First, second, and third nearest neighbors on a square lattice of the central site (in black) are denoted by $1$, $2$, and $3$, respectively. In the $k$-NN exclusion model, a particle excludes the sites up to the $k$th nearest neighbor from being occupied by another particle.}
	\label{fig:model}
\end{figure}

We consider a $L \times L$ square lattice or a $L \times L \times L$ cubic lattice with periodic boundary conditions. A lattice site may be empty or occupied by utmost one particle. In the $k$-NN exclusion model, a particle excludes all the sites up to the $k$th nearest neighbors from being occupied by another particle. Figure~\ref{fig:model}
shows the first, second, and third nearest neighbors on a square lattice. In the limit of large $k$, the model becomes equivalent to the problem of hard spheres in the continuum. In this paper, we study the $1$-NN, $2$-NN, and $3$-NN models in two and three dimensions.  These six models combined show a wide range of behavior: continuous transitions, first order transitions, multiple phase transitions, and columnar phase with sliding instability. Their phase diagram and nature of phase transitions are discussed in Sec.~\ref{sec:applications}.

For the application of the strip cluster update algorithm, certain lattice directions are preferred over the others (see Ref.~\cite{2014-nr-pre-multiple} for a more detailed discussion of this point). For all the models considered in this paper, except for the $3$-NN model in two dimensions, the preferred directions are the principal lattice directions. For the $3$-NN model in two dimensions, the preferred directions are along the $\pi/4$ and $-\pi/4$ diagonals~\cite{2014-nr-pre-multiple}. We refer to these preferred directions as rows. An important point is that, for all the models, along the rows, the minimum number of vacant sites between two particles is one.

We define the density $\rho$ to be $\rho=\eta/\eta_{max}$, where $\eta$ is the number density and $\eta_{max}$ is the number density of the fully packed phase. Thus the fully packed phase will always have $\rho=1$. The number of particles at full packing $N^{max}=\eta_{max} L^d$, where $d$ is the dimension. $\eta_{max}$ as well as the phase at fully packing for the different models studied in this paper are given in Table~\ref{tab:modelphi_max}. 
\begin{table}
\caption{\label{tab:modelphi_max}Number density at full packing, $\eta_{max}$, and the phase at full packing for the different models studied in the paper.  $d$ denotes the dimension.}
	\begin{ruledtabular}
		\begin{tabular}{c c c}
		Model	&  $\eta_{max}$ & Phase at full packing  \\
		\hline
		$1$-NN($2d$)&  $1/2$	& Sublattice\\
		$2$-NN($2d$)&  $1/4$	& Columnar\\
		$3$-NN($2d$)&  $1/5$	& Sublattice\\
		$1$-NN($3d$)&  $1/2$  & Sublattice\\
		$2$-NN($3d$)&  $1/4$ 	& Sublattice\\
		$3$-NN($3d$)&  $1/8$ 	& Columnar\\
	\end{tabular}	
	\end{ruledtabular}
\end{table}

\section{\label{sec:model b} Rejection-free cluster Wang-Landau algorithm}

In this section, we describe our main algorithm which we name as strip cluster Wang-Landau (SCWL) algorithm. Two main features of the algorithm are that it is based on cluster moves and that new configurations are weighted by their density of states. In order to establish the necessity of these two features for determining the density of states for generic HCLGs, we define two other algorithms for comparison: single site Wang-Landau (SSWL) algorithm based on single site moves and unbiased strip cluster Wang-Landau (USCWL) algorithm in which cluster moves are present but the new configurations that are generated are not weighted by their probabilities. These algorithms are described in Secs.~\ref{sec:sswl} (SSWL), \ref{sec:SCWL} (SCWL), and \ref{sec:USCWL} (USCWL).

First, we outline the WL protocol.
In the WL algorithm~\cite{2001-wl-prl-efficient,2001-wl-pre-determining}, a configuration with $N$ particles is weighted inversely proportional to $g(N)$, the number of configurations with $N$ particles. $g(N)$ changes continuously during the simulations and is expected to converge to its true value with increasing time. It is convenient to define the entropy 
\be
S(N) = \ln g(N).
\ee
Initially $S(N)=0$ for all $N$. The system is evolved using an evaporation-deposition algorithm that alters the number of particles consistent with their weights. A histogram $H(N)$ maintains the number of times configurations with $N$ particles are visited. After each evaporation-deposition move,  the entropy and histogram are updated as $S(N) \to S(N) +f$ and $H(N) \to H(N)+1$.  The system is evolved till the histogram becomes flat [$\min H(N)\geq c \max H(N)$], after which $f \to f/2$, and $H(N)=0$, and a new iteration is started. Here, $c$ is the predetermined constant that thresholds the ratio of the minimum to the maximum value of $H(N)$ for the flatness criterion to end an iteration. The iterations continue till $f$ reaches a predetermined small value.   Initially $f=1$; the value of $f$ is halved after each iteration. In our simulations, we choose $c=0.85$ and perform $22$ iterations so that the final value of $f$ is $2^{-22}$, unless otherwise specified. 

We now define three algorithms based on different evaporation-deposition moves.

\subsection{\label{sec:sswl}Single Site Wang Landau (SSWL)}

In SSWL implementation, the evaporation-deposition moves consist of updating single sites. Consider a configuration with $n_{old}$ particles. Pick a site at random. If occupied, remove the particle to obtain a new configuration with $n_{new}$ particles where $n_{new}=n_{old}-1$. If the site is empty, it is occupied with a particle, provided it does not violate the hard-core constraint. Then $n_{new}=n_{old}+1$ or $n_{new}=n_{old}$ depending on whether a particle is added or not. The new configuration is accepted with probability ${\rm min}\left[1,\frac{g(n_{old})}{g(n_{new})}\right]$.
 
After each step, the entropy and histogram are updated. $L^{d}$ updates correspond to one Monte Carlo time step.

\subsection{\label{sec:SCWL} Strip Cluster Wang Landau (SCWL)}

In one time step of SCWL, multiple particles are evaporated and deposited. The new configurations will be chosen proportional to their weights, making the implementation rejection free. The basic steps are described below.

First, choose a row at random. As mentioned in Sec.~\ref{sec:model}, a row refers to any of the principal directions for all the models except the $3$-NN model in two dimensions, for which a row refers to diagonals in the $\pm \pi/4$ directions. Imagine that all the particles in this row are removed. The row now breaks up into segments consisting of continuous empty sites separated by sites that are excluded from being occupied due to particles in neighboring rows. Note that there is the possibility of a segment being a ring due to periodic boundary conditions. 

Choose one of these segments at random and remove all the particles in it and reoccupy this segment with a new configuration that is chosen as follows. Let this segment have $\ell$ sites and let there be $N_0$ particles remaining in the system after removing particles from this segment. It is possible to occupy $0, 1, \ldots, n^\prime$ particles, where $n^\prime = [(\ell+1)/2]$ for a segment with open boundary conditions and $n^\prime = [\ell/2]$ for a segment with periodic boundary conditions. The refilling is done in two steps: first we determine the number of particles $n$ that should be deposited and second we choose a random configuration from all possible ways of placing $n$ particles in $\ell$ sites. The procedure is repeated till all segments are updated. The histogram and entropy are updated once all the segments in a row are updated.

Two aspects need to be quantified: how to determine $n$ and how to choose a random configuration (given $n$).

We define $C_o(\ell,n)$ as the number of ways $n$ particles can be placed on a segment of length $\ell$ with open boundary conditions. Likewise, $C_p(\ell,n)$ is the number of ways when the boundary conditions are periodic. 
We also define ${\rm Prob}_o(\ell,n)$ and ${\rm Prob}_p(\ell,n)$ as the probabilities of choosing $n$ particles for open and periodic boundary conditions, respectively. Then,
\bea
{\rm Prob}_o(\ell,n)&=&\frac{C_o(\ell,n)/g(N_0+n)}{\sum_{i=0}^{n^\prime} C_o(\ell,i)/g(N_0+i)}, \label{eqn:Prob_o}\\
{\rm Prob}_p(\ell,n)&=&\frac{C_p(\ell,n)/g(N_0+n)}{\sum_{i=0}^{n^\prime} C_p(\ell,i)/g(N_0+i)}.\label{eqn:Prob_p}
\eea
The combinatorial factors $C_o(\ell,n)$ and $C_p(\ell,n)$ for the $1$-NN model are given by (see the Appendix~\ref{sec:appendix} for derivation)
\bea
C_o(\ell,n)&=& \frac{(\ell-n+1)!}{(\ell-2 n+1)! n!},~n=0,1,\ldots, \left[\frac{\ell+1}{2} \right], \label{eqn:C_open}\\
C_p(\ell,n)&=& \frac{\ell (\ell-n-1)!}{(\ell-2 n)! n!},~n=0,1,\ldots, \left[\frac{\ell}{2} \right].
\label{eqn:C_per}
\eea
Equations~(\ref{eqn:Prob_o})--(\ref{eqn:C_per}) allow us to determine $n$.

Once $n$ is fixed, we need to specify how a random configuration with $n$ particles is chosen. Consider  an open segment. We start filling it from left to right. Consider the first site. The probability $p_o(\ell,n)$ of it being empty is
\be
p_o(\ell,n) = \frac{C_o(\ell-1,n)}{C_o(\ell,n)} = \frac{\ell-2n+1}{\ell-n+1}.
\label{eqn:p_o(l,n)}
\ee
If the first site is empty, we move to the next site, $\ell \to \ell-1$, $n$ remains the same, and the procedure is repeated. If the first site is occupied, we move to the next-nearest site, $\ell \to \ell-2$, $n \to n-1$ and the procedure is repeated.

For a ring, let $p_p(\ell,n)$ be the probability of the first site (any randomly chosen site) being empty. It is given by
\be
p_p(\ell,n) = \frac{C_o(\ell-1,n)}{C_p(\ell,n)} = \frac{\ell-n}{\ell}.
\label{eqn:p_p(l,n)}
\ee
If the first site is empty, we move to the next site, and the problem of occupation reduces to a problem of an open segment of length $\ell-1$ and $n$ particles.  If the first site is occupied, then it reduces to the problem of an open segment of length $\ell-3$ and $n-1$ particles.

Note that the factors $C_o(\ell,n)$,  $C_p(\ell,n)$, $p_o(\ell,n)$, and $p_p(\ell,n)$ do not depend on $g(n)$ and can be stored in the beginning of the program to save computing time.

A Monte Carlo time step corresponds to $2L$ row updates in two dimensions and $3L^{2}$ row updates in three dimensions.

\subsection{\label{sec:USCWL} Unbiased Strip Cluster Wang Landau (USCWL)}

In USCWL implementation, a row is updated segment by segment, like in SCWL. The difference with SCWL is that, in USCWL, we choose a configuration with equal probability from all possible configurations, while in SCWL these configurations are weighted differently according to the current $g(n)$. The implementation of the evaporation-deposition moves for USCWL is as follows. Choose a row at random and break it up into independent segments as defined previously in Sec.~\ref{sec:SCWL}. Let the segment be of length $\ell$. To generate a new configuration for the segment, first evaporate all the particles in the segment ($n_{old}$). The probability ${\rm Prob}_{o}(\ell,n_{new})$ of choosing a new configuration with $n_{new}$ particles is
\be
{\rm Prob}_{o}(\ell,n_{new}) = \frac{C_{o}(\ell,n_{new})}{\sum_{i=0}^{n^\prime}C_{o}(\ell,i)},
\ee
where $C_{o}(\ell,i)$, is the number of ways of occupying $\ell$ sites with $i$ particles, as given in Eq.~(\ref{eqn:C_open}). Once $n_{new}$ is decided, a random configuration consisting of $n_{new}$ particles is determined by following the procedure described in the paragraph following Eq.~(\ref{eqn:C_per}). The new configuration is accepted with probability ${\rm min}\left[1,\frac{g(n_{old})}{g(n_{new})}\right]$. For a segment with periodic boundary conditions, the procedure is similar.

The entropy and histogram are updated  after all of the segments in the row are refilled. One Monte Carlo move consists of updating $2L$ rows in two dimensions and $3L^{2}$ rows in three dimensions.

\section{\label{sec:comparisons}Comparing the Algorithms}

In this section, we compare the efficiency and effectiveness of the three algorithms---SCWL, SSWL, USCWL---defined in Sec.~\ref{sec:model}. We compare their performance for the $1$-NN and $2$-NN models in two dimensions. The entropy as well as the phase of the high density states differ in these models. For the $1$-NN model, there are only two fully packed configurations and the phase has sublattice order, while, in the $2$-NN model, the number of fully packed configurations increases exponentially with system size and the phase has columnar order.
Sections~\ref{sec:1NNAnalysis} and \ref{sec:2NNAnalysis} contain the analysis for the $1$-NN model and the $2$-NN models, respectively.

\subsection{\label{sec:1NNAnalysis} $1$-NN Model in two dimensions}

\begin{figure}
\centering
 \includegraphics[width= \columnwidth]{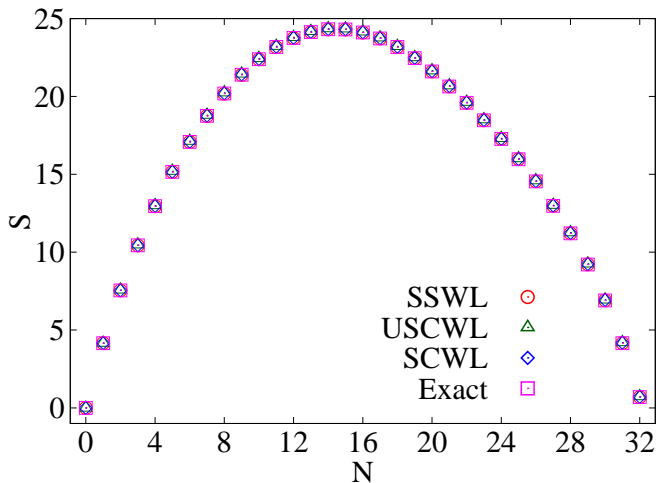}
\caption{(Color online) Entropy $S$ of the $1$-NN model in two dimensions for system size $L=8$, obtained from the algorithms SSWL, USCWL, and SCWL  at the end of $22$ iterations. The exact enumeration results are from Ref.~\cite{cunha2011critical}.} 
\label{fig:DOS1NN}
\end{figure}

We first benchmark our simulations by comparing the results for entropy $S(N)$ for $L=8$ obtained from the different algorithms with results from the exact enumeration. The arbitrariness in the zero of $S(N)$ is removed by setting $S(0)=0$. $S(N)$ for the different algorithms matches well with results from exact enumeration~\cite{cunha2011critical}, as shown in Fig.~\ref{fig:DOS1NN}. We conclude that the three different algorithms SSWL, USCWL, and SCWL sample the states correctly.

The error in the numerically estimated entropy is quantified by the error function $\epsilon$~\cite{belardinelli2007fast,2014-bpdl-jsm-intrinsic}: 
\begin{equation}
\epsilon  = \frac{1}{N^{max}-1}\sum_{N = 1}^{N^{max}}\left|1 - \frac{S(N)}{S_{ex}(N)}\right|,
\label{eqn:ErrorFunction}
\end{equation}
where $N^{max}=L^{2}/2$ is the maximal occupancy and $S_{ex}(N)$ is the exact entropy.
\begin{figure}
\centering
  \includegraphics[width=\columnwidth]{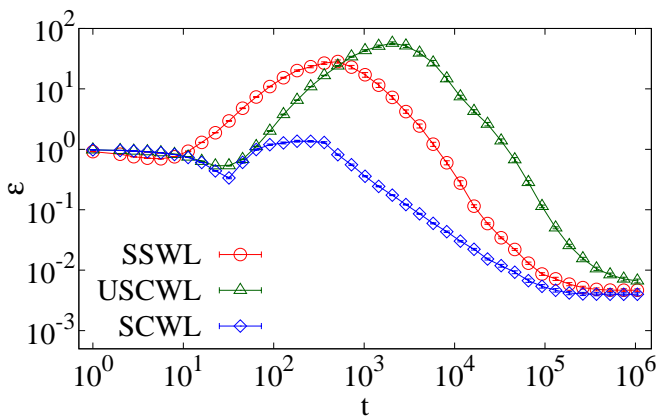}
\caption{
(Color online) Variation of the error function $\epsilon$ [see Eq.~(\ref{eqn:ErrorFunction})] with time $t$ for the $1$-NN model in two dimensions for system size $L=8$. For all three algorithms, the data have been averaged over $100$ realizations.
}
\label{fig:errorfn}
\end{figure}

The time dependence of $\epsilon$ for the three algorithms for $L=8$ is shown in Fig.~\ref{fig:errorfn}.  For both USCWL and SSWL, the error first increases significantly before decreasing to a time independent value. On the other hand, the error for SCWL is constant for initial times and then decreases to its final value. At all intermediate and large times, SCWL has a lower error, showing faster convergence. In addition, the saturation error is minimum for SCWL, showing better accuracy.

For larger system sizes, we do not know $S_{ex}$ and hence $\epsilon$ cannot be used as a measure for accuracy. In addition, $\epsilon$ does not tell us about how accurately the low-entropy states are accessed. In Table~\ref{tab:1nn2dalgo}, we compare the entropy of the states $N=1,2,N^{max}-1,N^{max}$ obtained using the three algorithms. $g(N)$ is easy to calculate exactly for these states. $g(1) = L^{2}, \; g(2) = L^{2}(L^{2}-5)/2, \; g(N^{max}-1) = L^{2}$, and $g(N^{max})=2$. For these values of $N$, the entropies are obtained for system sizes up to $L=36$. By examining the entropies for $N=N^{max}$ and $N=N^{max}-1$, it is clear that USCWL fails to estimate these entropies accurately. Also, the errors are the largest for USCWL. To compare convergence for larger $L$, we abort the routine if the time spent in any iteration exceeds $10^6$ Monte Carlo steps. Within this definition, USCWL fails to converge for $L \geq 24$. Both SCWL and SSWL give accurate estimates for entropies up to $L=24$. However, for $L=36$, SSWL fails to converge while SCWL continues to be accurate. We have checked that SSWL fails to converge even if we increase the cutoff for flattening of histogram to $10^{7}$ Monte Carlo steps. Also, we have checked that SCWL gives accurate results for these low entropy states even for $L=140$.
\begingroup
\squeezetable
\begin{table}
	\small
	\caption{
	Comparison of the entropy $S(N)$ for low entropy states obtained from SCWL, SSWL, and USCWL algorithms with the exact entropies. The data are for the $1$-NN model in two dimensions and have been averaged over $10$ realizations and after $20$ iterations. A blank space in any entry refers to cases where the time spent in an iteration exceeds  $10^{6}$ Monte Carlo steps without flattening the histogram. $N^{max}=L^{2}/2$ is the maximal occupancy. 
	}
  \begin{ruledtabular}
  \begin{tabular}{c c c c c}
	Algorithms & $S(1)$ & $S(2)$ & $S(N^{max}-1)$ & $S(N^{max})$  \\
	\hline
	\multicolumn{5}{c}{\begin{tabular}{c} \textbf{$L = 8$} \end{tabular}}\\
  %\hline
  SCWL & 4.160(8) & 7.546(8) & 4.14(2) & 0.68(2)\\
  SSWL & 4.158(7) & 7.546(6) & 4.16(2) & 0.69(2) \\
  USCWL & 4.16(2) & 7.57(2) & 4.20(2) & 0.73(2) \\
  %SANP & 4.19185 & 7.5823 & 4.21539 & 0.748616\\
  %\hline
  Exact & 4.159 & 7.543 & 4.159 & 0.693 \\
	%\hline
  \multicolumn{5}{c}{\begin{tabular}{c} \textbf{$L = 16$} \end{tabular}}\\
  %\hline
  SCWL & 5.549(3) & 10.376(6) & 5.52(2) & 0.68(2) \\
  SSWL & 5.549(5) & 10.382(6) & 5.56(4) & 0.69(4) \\
  USCWL & 5.66(9) & 10.58(9) & 5.8(1) & 1.0(1)\\
  %SANP & 5.76135 & 10.5009 & 5.6954 & 0.844667\\
  %\hline
  Exact & 5.545 & 10.378 & 5.545 & 0.693 \\
	%\hline
  \multicolumn{5}{c}{\begin{tabular}{c} \textbf{$L = 24$} \end{tabular}}\\
  %\hline
  SCWL & 6.357(5) & 12.011(6) & 6.33(2) & 0.67(2)\\
  SSWL & 6.364(7) & 12.02(1) & 6.33(6) & 0.64(6) \\
  USCWL &  &  &  &  \\
  %SANP & - & - & - & - \\
  %\hline
  Exact & 6.356 & 12.010  & 6.356 & 0.693 \\
	%\hline
	\multicolumn{5}{c}{\begin{tabular}{c} \textbf{$L = 36$} \end{tabular}}\\
  %\hline
  SCWL & 7.168(4) & 13.638(4) & 7.20(1) & 0.72(1)\\
  SSWL &  &  &  &  \\
  USCWL &  &  &  &  \\
  %SANP & - & - & - & - \\
  %\hline
  Exact & 7.167 & 13.637 & 7.167 & 0.693 \\
	%\hline
	\end{tabular}	
	\end{ruledtabular}	
  \label{tab:1nn2dalgo}
\end{table}
\endgroup

\begin{figure}
  \centering  
    \includegraphics[width= \columnwidth]{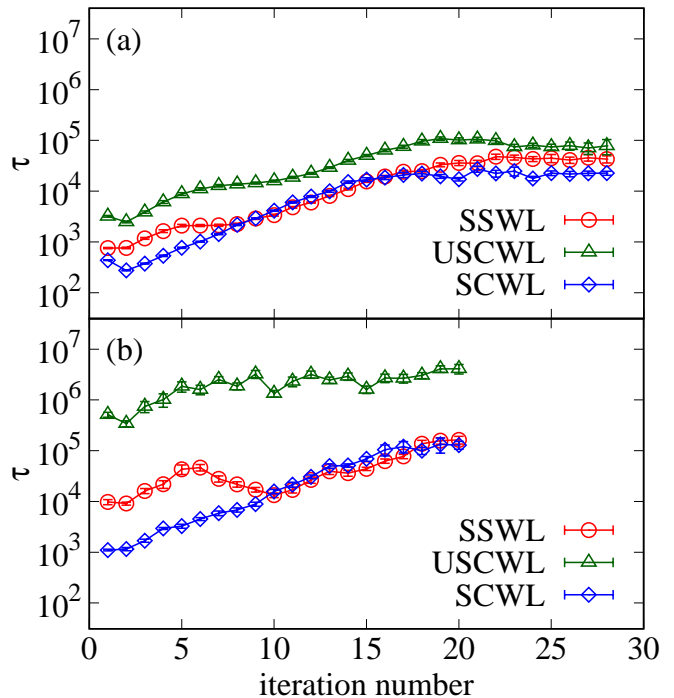}
    \caption{(Color online) Variation of $\tau$, the mean time taken for the histogram to flatten for a particular iteration, with iteration number for SSWL, USCWL, and SCWL   algorithms for the $1$-NN model in two dimensions. The data are for (a) $L=8$ and (b) $L=16$. The data have been averaged for $100$ realizations for $L=8$ and $10$ realizations for $L=16$.}
    \label{fig:ExitTimes}
\end{figure}

A measure of the rate of convergence is the time it takes to flatten the histogram in an iteration. We denote this time interval by $\tau$. Figure~\ref{fig:ExitTimes} shows the dependence of $\tau$ on iteration number for $L=8,\;16$ for the three algorithms. $\tau$ increases with iteration number and then saturates. It is clear that USCWL has a poor convergence rate compared to SSWL and SCWL. For the initial iterations, $\tau$ is much smaller for SCWL while, as the iteration number increases, SCWL and SSWL behave similarly.

We conclude, based on the data for $L=8,\;16,\;24,$ and $36$ for the $1$-NN model in two dimensions, that SCWL has the least error and fastest convergence. In addition, it is the only algorithm that is able to obtain results for $L \geq 36$ in reasonable computational time. We find that USCWL has poor performance compared to SCWL and SSWL on all parameters. We, therefore, do not use USCWL anymore. We make more detailed comparison between SCWL and SSWL in Sec.~\ref{sec:2NNAnalysis} for the $2$-NN model in two dimensions.

\subsection{\label{sec:2NNAnalysis}$2$-NN Model in two dimensions}

In this section, we further compare the performance of two algorithms, SSWL and SCWL, by using them to obtain the entropy for the $2$-NN model in two dimensions. Unlike the $1$-NN model, the degeneracy of the fully packed state in the $2$-NN model increases exponentially with system size. As a result, the sampling of the states near full packing becomes more challenging.

\begin{figure}
\centering
  \includegraphics[width= \columnwidth]{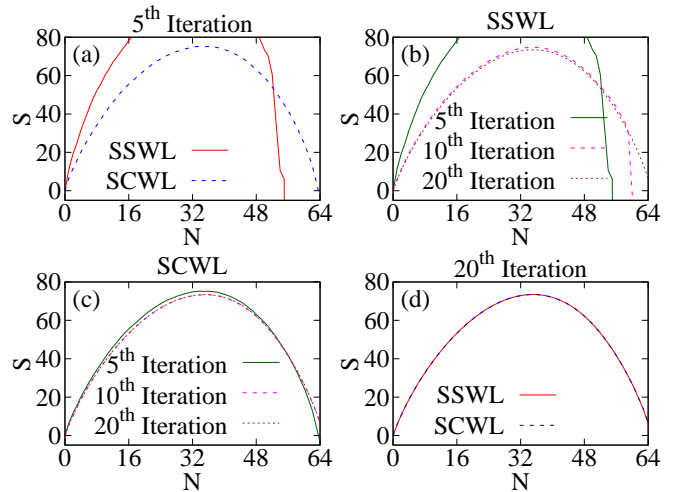}
  \caption{(Color online) Entropy $S$ at different iterations of the SSWL and SCWL algorithms for the $2$-NN model in two dimensions. The data are for $L=16$.  The different panels correspond to (a) $5$th iteration for SSWL and SCWL, (b) $5$th, $10$th, $20$th iteration for SSWL, (c) $5$th, $10$th, $20$th iteration for SCWL, and (d) $20$th iteration for SSWL and SCWL.}
  \label{fig:lnDOSK2}
\end{figure}

In Fig.~\ref{fig:lnDOSK2}, the entropy $S(N)$ at the $5$th, $10$th, and $20$th  iterations of the SSWL and SCWL algorithms is shown for $L=16$. $S(N)$ at the $5$th iteration obtained from the SSWL algorithm is significantly different from the final value [see Fig.~\ref{fig:lnDOSK2}(b)]. In this case as well as the $10$th iteration, the entropy is negative for states close to full packing, showing a slow convergence.  On the other hand, for the SCWL algorithm, the entropy at the $5$th and $10$th iterations are already close to the final result [see Fig.~\ref{fig:lnDOSK2}(c)]. The final entropies obtained from both algorithms are not distinguishable visually [see Fig.~\ref{fig:lnDOSK2}(d)].

To determine how well the algorithms sample the states near full packing, we compare the results from both algorithms with the exact entropy of the fully packed state. The latter can be computed to be $S(N^{max})= \ln[4 (2^{L/2}-1)]$. The percentage error in the numerically estimated entropy for $L=16$ is $9.52\%$ for the SSWL algorithm and $0.41\%$ for the SCWL algorithm. Clearly, the cluster moves employed in the SCWL algorithm considerably improve the accessibility of states near full packing, in addition to faster convergence.

For larger system sizes ($L=24$), we find that in the SSWL algorithm the histogram does not flatten within $10^{7}$ Monte Carlo steps. On the other hand, as we show in Sec.~\ref{sec:2nn2d}, we are able to obtain the density of states for $L$ up to $200$ using the SCWL algorithm.

\begin{figure}
\centering  
  \includegraphics[width= \columnwidth]{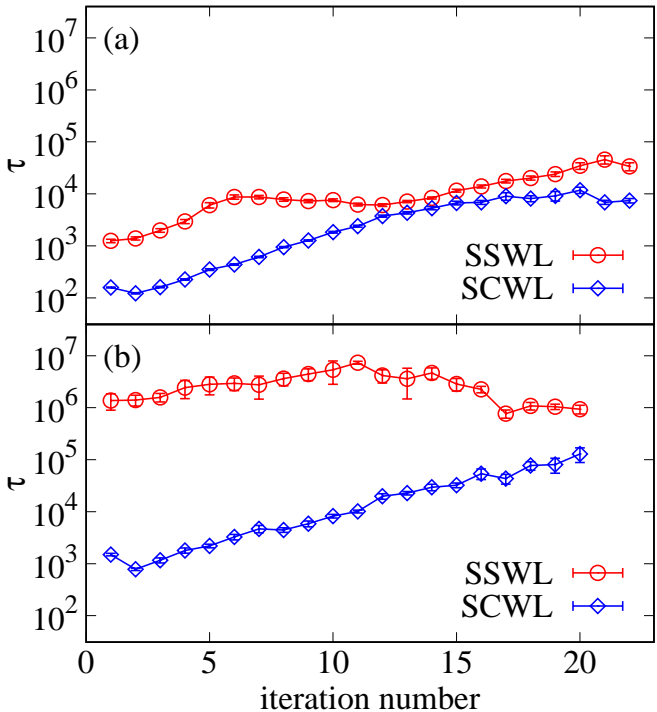}
  \caption{(Color online) Variation of $\tau$, the mean time taken for the histogram to flatten in a particular iteration, with iteration number for SSWL and SCWL algorithms for the $2$-NN model in two dimensions. The data are for (a) $L=8$ and (b) $L=16$. The data have been averaged for $100$ realizations for $L=8$ and $10$ realizations for $L=16$.}
  \label{fig:ExitTimesK2}
\end{figure}

We quantify the computational time by measuring $\tau$, the time it takes to flatten the histogram in an iteration. Figure~\ref{fig:ExitTimesK2} shows the dependence of $\tau$ on iteration number for $L=8,\;16$ for both the algorithms. For both $L$, $\tau$ for each iteration is larger for SSWL. The difference is enhanced with increasing $L$ with $\tau$ being nearly $1000$ times larger for SSWL for initial iterations for $L=16$. Also, SCWL takes much fewer iterations to reach the good estimate of $S(N)$, making it feasible to sample the density of states of much larger systems. 

We conclude, after comparing the performance of the algorithms for the $1$-NN and $2$-NN models in two dimensions, that both cluster moves as well as choosing new configurations proportional to their weights are essential to determine the density of states accurately.

\section{\label{sec:applications}Applications}
In this section, we show that the SCWL algorithm is efficient enough to accurately determine the critical behavior of the $k$-NN model for $k=1,2,3$  in two and three dimensions.
Knowing the density of states, we can calculate the average of any observable $O_{k,d}$:
\be
\langle O_{k,d} \rangle = 
\frac{\sum_{N=0}^{N^{max}} O(N) e^{\mu N}g(N)}{\sum_{N=0}^{N^{max}}e^{\mu N}g(N)},
\label{eqn:orderparameter}
\ee where $ N^{max}  $ is maximum occupancy for the $k$-NN model and $\mu$ is the chemical potential in units where $k_B T=1$, with $k_B$ being the Boltzmann constant and $T$ being the temperature. The subscripts $k$ and $d$ denote the range of exclusion and spatial dimension, respectively.

It is convenient to fix the notation for all the models in one place. We will denote the (intensive) order parameter by $q_{k,d}$. The definition of $q_{k,d}$ will depend on the symmetries of the model. The other thermodynamic quantities that we will be interested in are the  compressibility $\kappa_{k,d}$, susceptibility $\chi_{k,d}$, and pressure $P_{k,d}(\mu)$, which are defined as
\bea
\kappa_{k,d}&=& L^{d} \left(\langle \rho_{k,d}^{2} \rangle - \langle \rho_{k,d} \rangle^2 \right),
\label{eq:kappa}\\
\chi_{k,d} &=& L^{d} \left(\langle q_{k,d}^{2} \rangle-\langle q_{k,d} \rangle^{2} \right),
\label{eq:chi}\\
P_{k,d}(\mu) &=& L^{-d} \ln \sum_{n = 0}^{N^{max}}e^{\mu n} g(n).
\label{eq:P}
\eea
We can also measure pressure in the canonical ensemble,  $ \widetilde{P}_{k,d}$:
	\begin{equation}
\widetilde{P}_{k,d} (\rho) =\int_{0}^{\rho} (1-\phi(\rho)) \frac{\partial}{\partial \rho} \left[ \frac{\rho}{1-\phi(\rho)}\right] d \rho,\label{eq:Pphi}
	\end{equation} 
where $\phi(\rho)$ is the mean fraction of sites that are blocked from further occupation at density $\rho$~\cite{darjani2017extracting,darjani2019liquid}.  $\phi(\rho)$ is directly measured in the flat histogram algorithm, allowing  $\widetilde{P}$ to be calculated.  Finally, we will denote the density by $\rho_{k,d}$ or by just $\rho$ if there is no cause for confusion.

Phase transitions are characterized by the nonanalytic behavior of the thermodynamic quantities, which is captured by the critical exponents~\cite{fisher1967theory}. In finite systems, the behavior gets rounded off, but can be captured through finite size scaling~\cite{fisher1972scaling,fisher1983scaling,pelissetto2002critical}. Near a continuous transition, the finite size scaling of the different quantities are 
\be 
\begin{split}
\kappa_{k,d} &\approx L^{\alpha/\nu} f_\kappa\left(\epsilon L^{1/\nu}\right),  \\
\langle q_{k,d} \rangle &\approx L^{-\beta/\nu} f_q\left(\epsilon L^{1/\nu}\right), \\
\chi_{k,d} &\approx L^{\gamma/\nu} f_\chi\left(\epsilon L^{1/\nu}\right),
\end{split}
\label{eq:fullscaling}
\ee
where $\alpha$, $\beta$, $\gamma$, and $\nu$ are critical exponents, $\epsilon=\mu-\mu_c$ is the deviation from the critical point, and $f$ are scaling functions. At a first order transition, similar scaling behavior is seen with $\nu=1/d$ and $\alpha/\nu=\beta/\nu=\gamma/\nu=d$. 

For the numerical analysis, it is useful to define an associated quantity, which we will denote by $t$:
\be
t_{k,d}=\frac{\partial \ln \langle q_{k,d} \rangle}{\partial \mu}.
\label{eq:t}
\ee
From Eq.~(\ref{eq:fullscaling}), we obtain
\be
t_{k,d} \approx L^{1/\nu} f_t\left(\epsilon L^{1/\nu}\right).
\label{eq:tscaling}
\ee
The advantage of using $t_{k,d}$ is that the maxima scale as $L^{1/\nu}$, allowing for a single parameter determination of $\nu$. 

To determine the critical parameters at the continuous transitions, we  determine exponents one at a time. $1/\nu$, $\gamma/\nu$, and $\alpha/\nu$ are determined from the scaling of $t$, $\chi$, and $\kappa$, respectively, with $L$. $\mu_c(L)$ is identified with the position of the  peak of $\chi$.  $-\beta/\nu$ is obtained by the scaling of $q(\mu_c(L))$ with $L$. These power-law scalings are summarized as
\be
\begin{split}
\chi_{k,d}^{\mathrm{max}} &\sim L^{\gamma/\nu},\\
t_{k,d}^{\mathrm{max}} &\sim L^{1/\nu}, \\
\kappa_{k,d}^{\mathrm{max}} &\sim L^{\alpha/\nu}, \\
\langle q_{k,d} \rangle(\mu_c) &\sim L^{-\beta/\nu}.
\label{eq:maxscaling}
\end{split}
\ee 
Let $\mu_c$ and $\rho_c$ denote the critical chemical potential and critical density in the thermodynamic limit. The system size dependent critical chemical potential and density are extrapolated to the infinite system size limit using
\bea
\mu_{c}(L)  -  \mu_{c} &\sim  L^{-1/\nu}, \label{eq:muscaling} \\
\rho_{c}(L)  -  \rho_{c} &\sim  L^{-1/\nu}. \label{eq:rhoscaling} 
\eea

To estimate critical exponents accurately, the relevant thermodynamic quantities were estimated with a step size of $\Delta \mu = 10^{-5}.$ Errors in each data point are standard error obtained from 16 independent simulations using different sequences of random numbers.  Errors in the final estimate of critical parameters are fitting errors.

\subsection{$1$-NN Model in two dimensions} \label{sec:1nn2d}
In the $1$-NN model, the four nearest neighbors of a particle are excluded from being occupied. As density is increased, the system is known to undergo a continuous transition from a disordered fluid phase to an ordered sublattice phase (see~\cite{2007-fal-jcp-monte,rodrigues2021husimi} and references within for the large body of work on this model). The transition is expected to belong to the Ising universality class: $\gamma /\nu=7/4$, $\beta/\nu=1/8$, $\alpha/\nu=0$, and $\nu=1$~\cite{2007-fal-jcp-monte,2011-dc-pre-complete}. The best known numerical estimates of the critical chemical potential and critical density, obtained from transfer matrix calculations, are $\mu_{c,1,2d}=1.33401510027774(1)$ and  $\rho_{c,1,2d}=0.7354859980820(6)$~\cite{2002-gb-pre-finite} (note that the density $\rho$ is two times the number density). 

\begin{figure}
	\includegraphics[width=0.8\columnwidth]{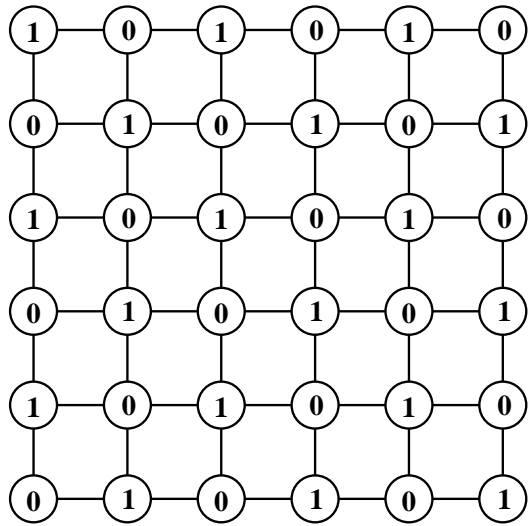}
	\caption{For the $1$-NN model in two dimensions, the square lattice is  divided into two sublattices labeled by $0$ and $1$.}
	\label{fig:2d1NNsublattice}
\end{figure}

To define the order parameter, we divide the square lattices into two sublattices as shown in Fig.~\ref{fig:2d1NNsublattice}. At full packing, only  one of the sublattices is occupied. The order parameter is defined as
\be
\langle q_{1,2d} \rangle = \left| \rho_1 -\rho_0 \right|,
\label{eqn:OP1NN}
\ee
where $\rho_{i}$ is the density of particles in sublattice $i$. In the fluid phase $\langle q_{1,2d} \rangle$ is zero and in the sublattice phase $\langle q_{1,2d} \rangle$ is nonzero. 

\begin{figure}
\centering
 \includegraphics[width= \columnwidth]{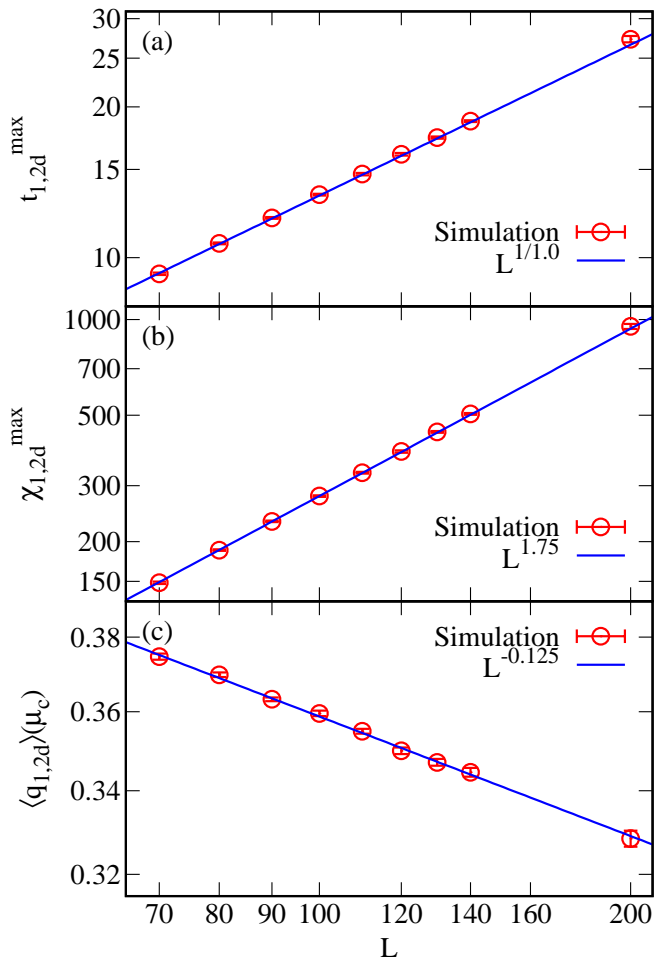}
\caption{(Color online) Power-law fits for the scaling of (a) $t_{1,2d}^{\mathrm{max}}$, (b) $\chi_{1,2d}^{\mathrm{max}}$, and (c) $\langle q_{1,2d} \rangle (\mu_c)$ with system size $L$ for the $1$-NN model in two dimensions.  The axes are scaled logarithmically.} 
\label{fig:fitcurves}
\end{figure}

\begin{figure}
\centering
 \includegraphics[width= \columnwidth]{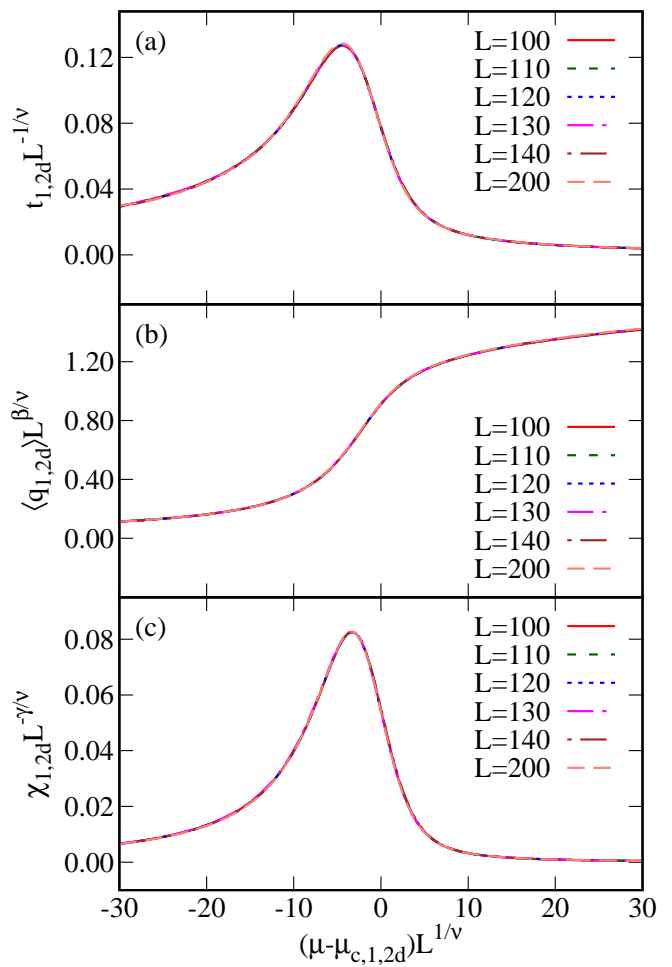}
\caption{(Color online) Data for the $1$-NN model in two dimensions for  different system sizes collapse onto one curve for (a) $t_{1,2d}$, (b) $\langle q_{1,2d} \rangle$, and (c) $\chi_{1,2d}$, when scaled as in Eq.~(\ref{eq:fullscaling}) with exponents $\nu = 1.00(1)$, $\beta/\nu=0.125(4)$, $\gamma/\nu=1.75(1)$, and $\mu_{c,1,2d} = 1.3345(6)$.}
\label{fig:CriticalExponents1NN}
\end{figure}

 We determine the density of states for system sizes up to $ L=140 $.
We determine the critical exponents using Eq.~(\ref{eq:maxscaling}). The power-law scaling and the best fits are shown in Fig.~\ref{fig:fitcurves} for $t_{1,2d}^{\mathrm{max}}$,  $\chi_{1,2d}^{\mathrm{max}}$, and $q_{1,2d}(\mu_c(L))$. We obtain $\nu=1.00(1)$, $\gamma/\nu=1.75(1)$ and $\beta/\nu=0.125(4)$.  We have also shown the data for $L = 200$ in Fig.~\ref{fig:fitcurves}, which falls on the same line as obtained by data fit for sizes up to $L=140$.  Extrapolating $\mu_c(L)$ and $\rho_c(L)$ using Eqs.~(\ref{eq:muscaling}) and (\ref{eq:rhoscaling}), we obtain 
$\mu_{c,1,2d}= 1.3345(6)$ and $\rho_{c,1,2d}= 0.7332(6)$. The estimate for $\mu_{c,1,2d}$ is  consistent with known estimates (see above). The critical density  $\rho_{c,1,2d}$ differs from the best known estimate by $0.3\%$.  The data for the thermodynamic quantities for different system sizes collapse onto one curve when scaled as in Eq.~(\ref{eq:fullscaling}) with the numerically obtained critical parameters (see Fig.~\ref{fig:CriticalExponents1NN}).

\subsection{$2$-NN Model in two dimensions} \label{sec:2nn2d}
In the $2$-NN model in two dimensions, a particle excludes eight sites from being occupied by another particle. It is known that the system undergoes a continuous phase transition from a low density disordered phase to a high density columnar phase. In the columnar phase, particles preferentially occupy either even or odd rows with no preference for the parity of columns or even or odd columns with no preference for parity of rows. 

The  disordered-columnar transition belongs to the Ashkin-Teller  universality class~\cite{2015-rdd-prl-columnar}. The Ashkin-Teller model has a line of critical points. Along this line $\gamma/\nu$ and $\beta/\nu$ are constant and equal $\gamma/\nu=7/4$ and $\beta/\nu=1/8$.  The critical line is parametrized by the exponent $\nu$. For the $2$-NN model, it has proved difficult to obtain precise estimates of $\nu$. More recent estimates have been $\nu=0.92(3)$~\cite{2015-rdd-prl-columnar} from transfer matrix based Monte Carlo simulations,  $\nu=0.86(2)$~\cite{2007-zt-prb-lattice} from exchange Monte Carlo method, $\nu=0.94(3)$~\cite{2011-fbn-pre-lattice} from  Monte Carlo simulation, and $\nu= 1.0$~\cite{2007-fal-jcp-monte} from Monte Carlo simulations. The known estimates for critical chemical potentials are  $\mu_{c,2,2d}=4.58(4)$~\cite{2015-rdd-prl-columnar}  from transfer matrix based Monte Carlo simulations,   $\mu_{c,2,2d}=4.56(2)$~\cite{2007-zt-prb-lattice} from exchange Monte Carlo method, $\mu_{c,2,2d}=4.584(2)$~\cite{2011-fbn-pre-lattice} from  Monte Carlo simulation and $\mu_{c,2,2d}=4.578$~\cite{2007-fal-jcp-monte} from Monte Carlo simulations.  The corresponding estimates for critical density are $\rho_{c,2,2d}=0.96$~\cite{1984-akg-prb-square},  $0.932$~\cite{2007-fal-jcp-monte}, and  $0.930(1)$~\cite{2007-zt-prb-lattice}. The intractability of the model has resulted in many attempts to obtain the critical density and chemical potential using systematic expansions and approximate methods. These include high activity expansions~\cite{1967-bn-jcp-phase,2012-rd-pre-high,2015-nkr-jsp-high}, estimates of surface tension between ordered phases~\cite{1983-s-jpc-phase,2016-ndr-epl-stability,mandal2017estimating}, and limits of Husimi tree~\cite{rodrigues2021husimi}. 

The order parameter is defined as
\be
q_{2,2d} = \sqrt{(\rho_{oc}-\rho_{ec})^2 +(\rho_{or}-\rho_{er})^2 }\label{eqn:OP2NN}, 
\ee
where the subscripts $o,e,r,c$ denote odd, even, row, and column, respectively.  $\rho_{oc}$ is the density of particles in odd columns and so on. $q_{2,2d}$ becomes nonzero when the odd-even parity is broken.

\begin{figure}
	\includegraphics[width=\columnwidth]{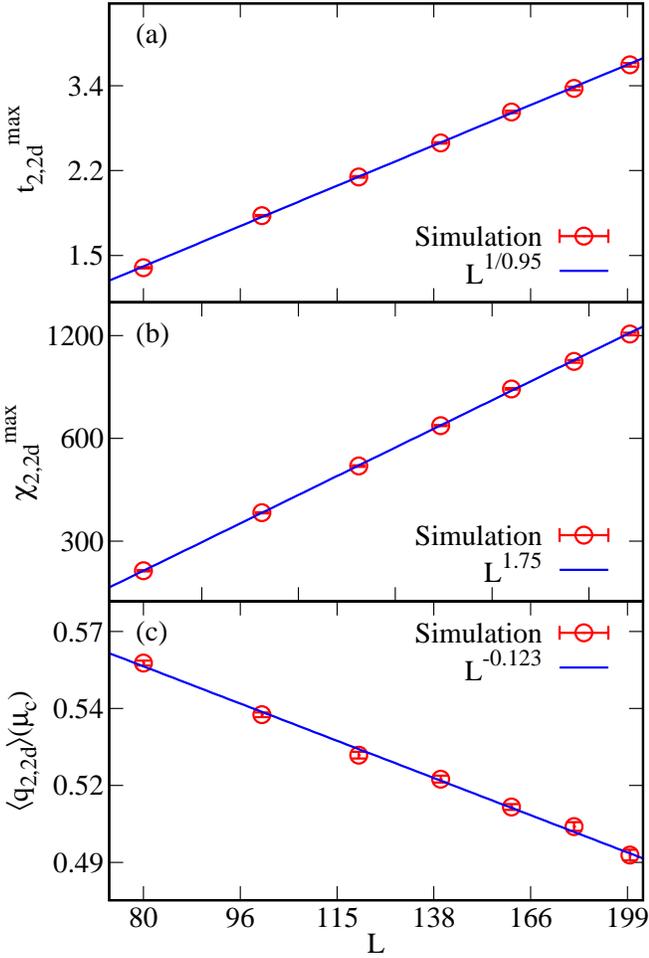}
	\caption{ (Color online) Power-law fits for the scaling of (a) $t_{2,2d}^{\mathrm{max}}$, (b) $\chi_{2,2d}^{\mathrm{max}}$, and (c) $\langle q_{2,2d} \rangle (\mu_c)$ with system size $L$  for the $2$-NN model in two dimensions. The axes are scaled logarithmically.}
	\label{fig:2d2NNnugamma}
\end{figure}

We determine the density of states  for system sizes up to $ L=200 $. From the scaling of $\chi_{2, 2d}^{\mathrm{max}}$ and $q_{2, 2d}(\mu_c(L))$ [see Fig.~\ref{fig:2d2NNnugamma}(b) and (c)], we obtain $\gamma/\nu=1.75(1)$ and $\beta/\nu=0.123(3)$. Both these estimates are consistent with the Ashkin-Teller values $\gamma/\nu=1.75$ and $\beta/\nu=0.125$. From the scaling of $t_{2, 2d}^{\mathrm{max}}$ [see Fig.~\ref{fig:2d2NNnugamma}(a)], we obtain $\nu=0.95(2)$.  This estimate is consistent with recent estimates of $\nu$ (see second paragraph of this subsection). We note that these estimates are with using system sizes only up to $L=200$. By using more sophisticated methods like flat histogram with windows, etc., it would be possible to study much larger system sizes. This in turn should result in even better estimates of $\nu$.

\begin{figure}
	\includegraphics[width=\columnwidth]{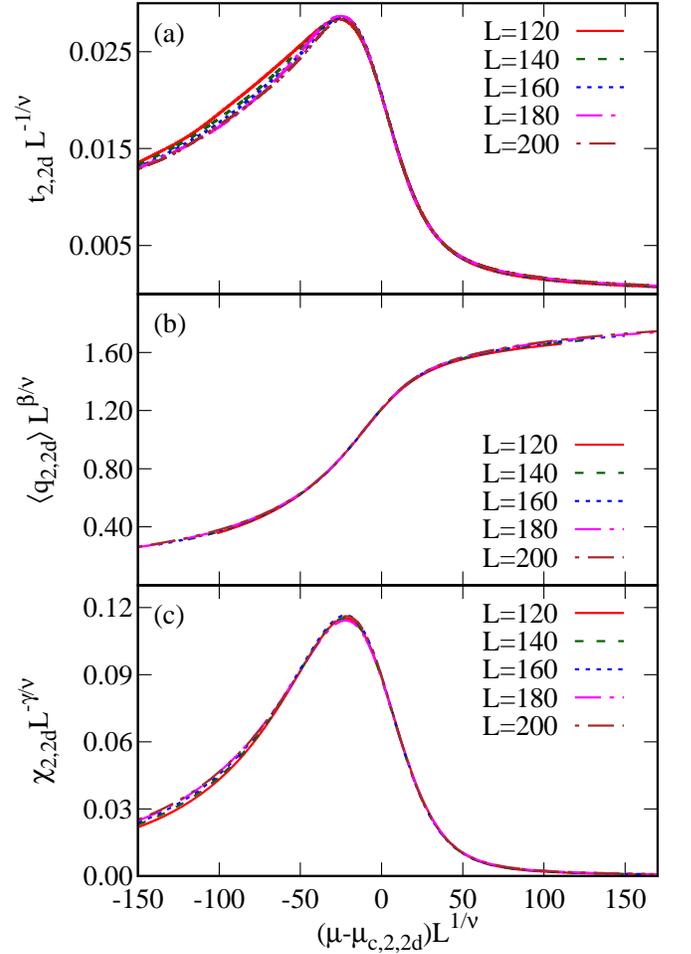}
	\caption{(Color online) Data for the $2$-NN model in two dimensions for different system sizes collapse for (a) $t_{2,2d}$, (b) $\langle q_{2,2d} \rangle$, and (c) $\chi_{2,2d}$, when scaled as in Eq.~(\ref{eq:fullscaling}) with exponents $\nu = 0.95(2)$, $\beta/\nu=0.123(3) $, $\gamma/\nu=1.75(1)$, and $\mu_{c,2,2d} = 4.580(4)$.}	\label{fig:2d2NNcollapse}
\end{figure}

To find critical parameters we extrapolate $\mu_{c}(L)$ and $\rho_c(L)$ to infinite system size using Eqs.~(\ref{eq:muscaling}) and (\ref{eq:rhoscaling}). We obtain $\mu_{c,2,2d}=4.580(4)$. This value agrees very well with best earlier estimate $4.58(4)$.  We also obtain $\rho_{c,2,2d}=0.9307(3)$,  again consistent with earlier estimates. The data for the thermodynamics quantities for different system sizes collapse onto one curve when scaled as in Eq.~(\ref{eq:fullscaling}) with the numerically obtained critical parameters and exponents (see Fig.~\ref{fig:2d2NNcollapse}).

\subsection{\label{sec:3nn2d}$3$-NN Model in two dimensions}
In the $3$-NN model in two dimensions, a particle excludes $12$ sites from being occupied by another particle. It is known that the system undergoes a discontinuous phase transition from a low density disordered fluid phase to a high density sublattice ordered  phase. The known estimates for critical chemical potential are 
$\mu_{c,3,2d}=3.6758(8)$~\cite{2013-fl-jcp-exploiting} and $3.6762(1)$~\cite{2005-eb-epl-first}. At the first order transition, the known estimates for the coexistence densities $\rho_{f}$ and $\rho_{s}$, where $f$ and $s$ denote fluid and solid, are $\rho_{f}=0.80$ and $\rho_{s}=0.95$~\cite{1967-bn-jcp-phase,1982-ob-jpa-phase,2005-eb-epl-first,2000-eb-jpa-random,rotman2009ideal,rotman2010direct} (note that the density $\rho$ is five times the number density). The value of critical pressure has been estimated to be $0.74124(2)$ from the matrix method~\cite{2005-eb-epl-first} and $0.74147(2)$ from high density series expansion~\cite{2005-eb-epl-first}.
\begin{figure}
	\includegraphics[width=\columnwidth]{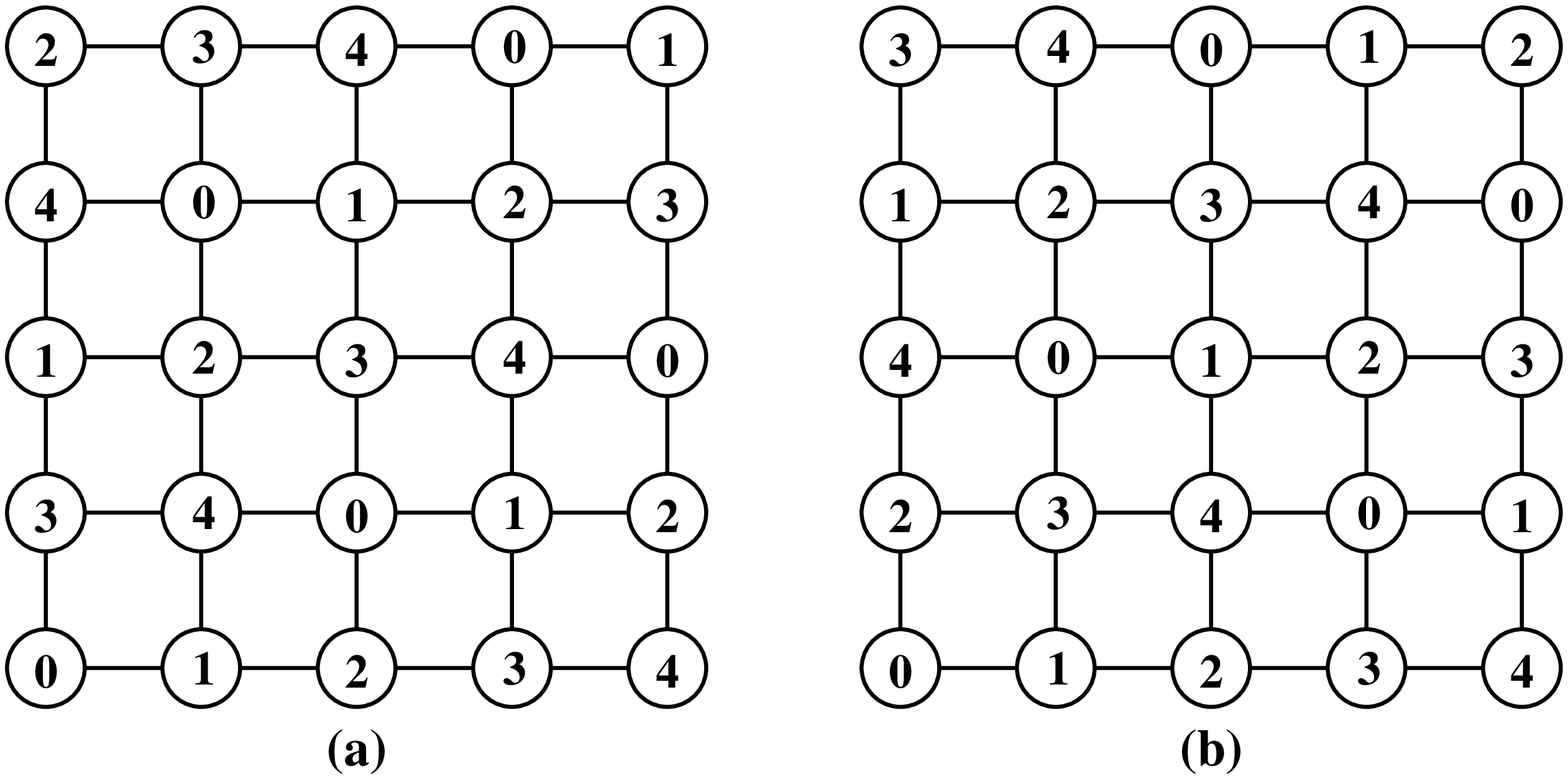}
	\caption{For the $3$-NN model in two dimensions, the square lattice is  divided into two sublattices labeled by $0$-$4$. Two divisions are possible, which are denoted as (a) type $A$ and (b) type $B$.
}
	\label{fig:2d3NNsublattice}
\end{figure}

To define the order parameter, we divide the lattice sites into five sublattices as shown in Fig.~\ref{fig:2d3NNsublattice}. This division can be done in two ways, which we call type-$A$ or type-$B$ sublattices.  At full packing, one of the sublattices of either type $A$ and type $B$ are fully occupied, and in the disordered phase all five sublattices of both types are equally occupied on an average. Let
\be
  q_{p} = \left |\sum_{i=0}^{4} \rho_{i}^{p}\exp\left[\; j\frac{2\pi i}{5}\right]\right |,~~p=A, B,
 \label{eqn:OP3NN2}
 \ee
where $\rho_{i}^{p}$ is the number density of particles in sublattice $i$ of type $p$. $q_p$ is nonzero when a particular sublattice of type $p$ is preferred.
We define the  order parameter to be
\be
\langle q_{3,2d} \rangle = \left | q_{A}-q_{B}\right |.
\label{eqn:OP3NN1}
\ee

\begin{figure}
\centering
  \includegraphics[width= \columnwidth]{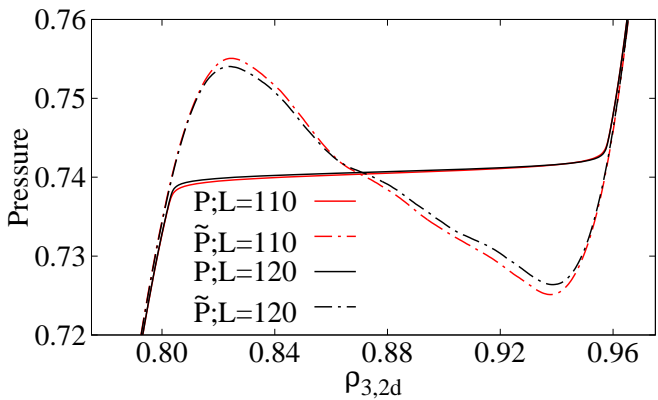}
\caption{(Color online)  Variation of the grand canonical pressure $P$ computed from Eq.~(\ref{eq:P}) and the canonical pressure $\widetilde{P}$ computed from Eq.~(\ref{eq:Pphi}), with density for the $3$-NN model in two dimensions. The data are for the two largest system sizes studied.}
\label{fig:Pressure3NN}
\end{figure}

\begin{figure}
\centering
  \includegraphics[width= \columnwidth]{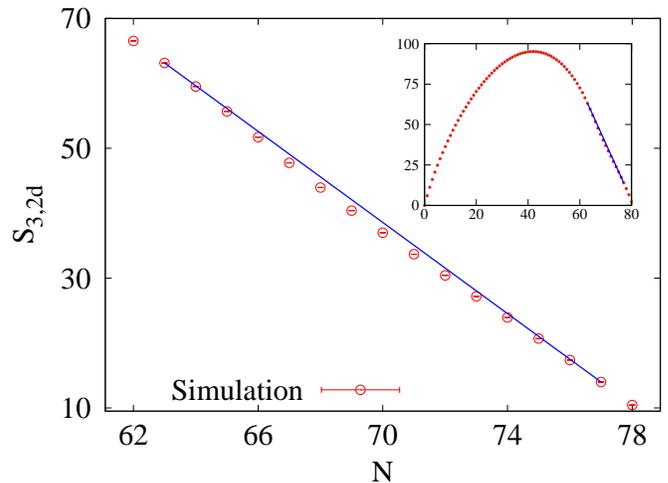}
\caption{(Color online) Nonconvex nature of entropy of the $3$-NN model in two dimensions. The solid straight line is a convex envelope with points on this line having higher entropy than the measured value. The data are for $L=20$. Inset shows full range of entropy and position of the convex envelope.}
\label{fig:Entropy3NN}
\end{figure}

We determine the density of states for system sizes up to $L=120$. The first order nature of the transition can be established by studying the pressure and entropy.  Figure~\ref{fig:Pressure3NN} shows the variation of pressure with density, computed both in the grand canonical ensemble ($P$) as well as the canonical ensemble ($\widetilde{P}$).  $\widetilde{P}$ is nonmonotonic, while $P$ is nearly a constant in the coexistence regime. The loops in $\widetilde{P}$ are possibly due to a finite size effect caused by the interface between a bubble of minority phase and the surrounding majority phase~\cite{2011-bk-prl-twostep}. The curve for $P$ is similar to the usual Maxwell construction for a nonmonotonic $\widetilde{P}$.
The pressure loop in $\widetilde{P}$ would imply nonconvexity in the entropy. The nonconvexity of the entropy is demonstrated in Fig.~\ref{fig:Entropy3NN}. As can be seen, entropy is convex everywhere except in a small interval covered by the convex envelope (straight line in Fig.~\ref{fig:Entropy3NN}) where the measured entropy is lower than the entropy of a phase separated system. This feature persists for all system sizes that we have studied.

From the convex envelope construction, the critical parameters can be accurately measured. We identify the end points of the convex envelope with the coexistence densities $\rho_{f}(L)$ and  $\rho_s(L)$. The critical chemical potential is given by
\begin{equation}
\mu_c(L)=-\frac{S(\rho_s)-S(\rho_f)}{N_s-N_f}.  \label{eqn:NC}
\end{equation}
$\rho_f(L)$, $\rho_s(L)$, and $\mu_c(L)$, obtained from both convex envelope as well as the peak of susceptibility, are tabulated in Table~\ref{tab:extn_nconc3NN} for different system sizes. The estimates for $\mu_c(L)$ obtained from both methods are very close to each other.
\begin{table}
	\caption{\label{tab:extn_nconc3NN} Critical  parameters obtained from  nonconvexity (NC) of entropy  for the $3$-NN model in two dimensions. The data are extrapolated to infinite system size using  linear regression with $ L^{-2} $. 
}
	\begin{ruledtabular}
		\begin{tabular}{c c c c c}
 $L$ & $\rho_{f}(L)$ & $\rho_{s}(L)$ & $\mu_{c,3,2d}(L)$ &   $\mu_{c,3,2d}(L)$ \\
			&  			   & 			 & from NC & from $\chi^{max}_{3,2d}$ \\
		\hline
		%20	& 0.79(1)		& 0.96(1)	&	3.5112(1) & 3.4994(2)\\
		%30 	& 0.794(6)		& 0.961(6)	&	3.6000(3) & 3.5959(3)\\
		%40	& 0.797(3)		& 0.960(3)	&	3.6327(3) & 3.6308(3)\\
		%50	& 0.800(2) 		& 0.960(2)	&	3.6479(2) & 3.6470(2)\\
		60	& 0.801(1) 		& 0.959(1)	&	3.6549(1) & 3.6544(1)\\
		70	& 0.802(1)		& 0.958(1)	&	3.6615(1) & 3.6613(1)\\
		80 	& 0.8024(8)		& 0.9581(8)	&	3.6648(1) & 3.6647(1)\\
		90	& 0.8032(6)		& 0.9580(6)	&	3.6672(1) & 3.6671(1)\\
		100	& 0.8035(5)  	& 0.9577(5)	&	3.6688(1) & 3.6688(1)\\
		110	& 0.8040(4) 	& 0.9576(4)	&	3.6700(1) & 3.6700(1)\\
		120	& 0.8042(4) 	& 0.9575(4)	&	3.6712(1) & 3.6712(1)\\
		%\hline \hline
		$\infty$ & 0.8055(3)& 0.9570(3)	&	3.6766(5) & 3.6764(6)\\
	\end{tabular}	
	\end{ruledtabular}
\end{table}

\begin{figure}
	\centering
	\includegraphics[width= \columnwidth]{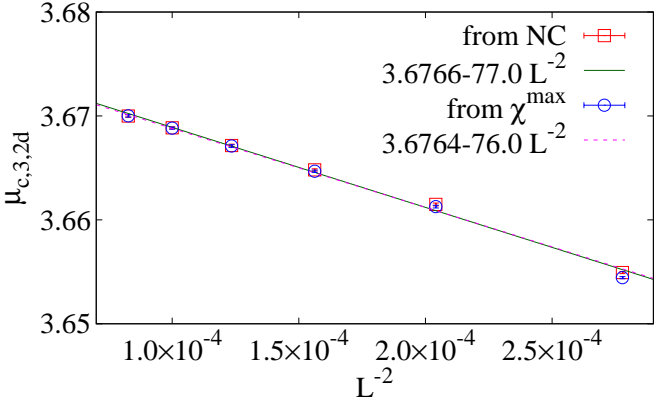}
	\caption{(Color online)  Extrapolation of  critical chemical potential $\mu_{c,3,2d}(L)$ to infinite system size for the 3-NN model in two dimensions.} 	
	\label{fig:fitcurves3NN}
\end{figure}

We extrapolate the critical parameters to infinite system size using Eqs.~(\ref{eq:muscaling}) and (\ref{eq:rhoscaling}) with $\nu=1/2$. As an example, we show the  extrapolation for  $\mu_{c}(L)$ in Fig.~\ref{fig:fitcurves3NN}.
We obtain $\mu_{c,3,2d}=3.6766(5)$ from the nonconvex analysis and $\mu_{c,3,2d}=3.6764(6)$ from the analysis of susceptibility. These values are close to earlier estimates of  $3.6758(8)$~\cite{2013-fl-jcp-exploiting} and $3.6762(1)$~\cite{2005-eb-epl-first}. For the coexistence densities, we obtain $\rho_{f,3,2d}=0.8055(3)$ and $\rho_{s,3,2d}=0.9570(3)$.  This improves the earlier estimates of $\rho_{f}=0.80$ and $\rho_{s}=0.95$~\cite{1967-bn-jcp-phase,1982-ob-jpa-phase,2005-eb-epl-first,2000-eb-jpa-random,rotman2009ideal,rotman2010direct}.  To obtain the system size dependent critical pressure, we determine the pressure at $\mu_c(L)$. Extrapolating to infinite system size, we obtain the critical pressure to be $P_{c,3,2d}=0.74147(6)$. This equals earlier estimates from  high density series expansion~\cite{2005-eb-epl-first}.

\begin{figure}
	\centering
	\includegraphics[width= \columnwidth]{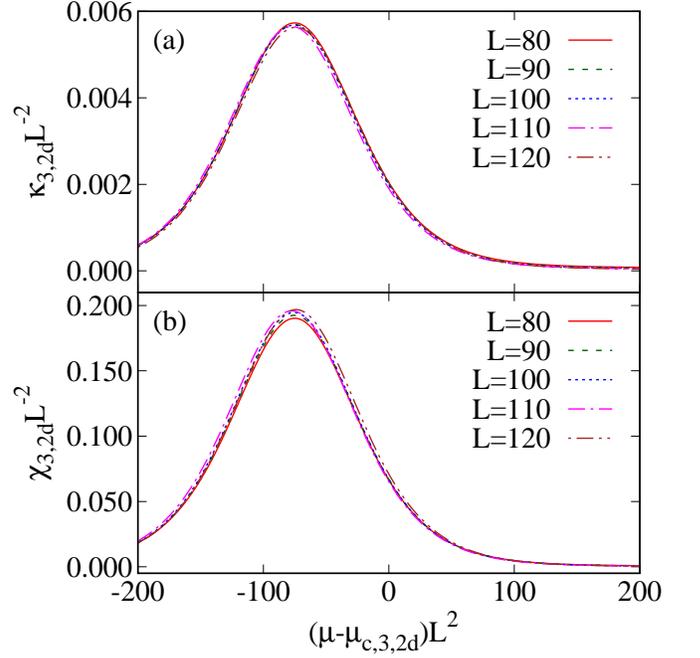}
	\caption{ (Color online) Data for different system sizes collapse for (a) $\kappa_{3,2d}$ and  (b) $\chi_{3,2d}$, when scaled as in Eq.~(\ref{eq:fullscaling}) with $\mu_{c,3,2d} = 3.6764(6)$ and $\nu=1/d$ (for first order transition). The data are for the  $3$-NN model in two dimensions.
}	\label{fig:ScalingCollapse3NN}
\end{figure}

Finally, we show that the data for susceptibility and compressibility for different system sizes collapse onto one curve when scaled as in Eq.~(\ref{eq:fullscaling}) with the numerically obtained critical parameters and the exponents for a first order transition (see Fig.~\ref{fig:ScalingCollapse3NN}).

\subsection{\label{sec:1nn3d}$1$-NN Model in three dimensions}
In the $1$-NN model in three dimensions, a  particle excludes six nearest neighbor sites from being occupied by another particle. The system undergoes a single  continuous  phase transition from a low density disordered phase to a high density sublattice phase when density is increased~\cite{1996-hb-physica-simple,2005-p-jcp-thermodynamic,cunha2011critical}. From symmetry considerations, the transition is expected to belong to the three dimensional  Ising universality class. Earlier estimates of the critical value of the chemical potential are $\mu_{c,1,3d}=0.05443(7)$~\cite{1996-hb-physica-simple}, $0.0503(100)$~\cite{2005-p-jcp-thermodynamic}, and $0.0552(7)$~\cite{cunha2011critical}, while that of the critical density is  $\rho_{c,1,3d}=0.42164(10)$~\cite{cunha2011critical}. The known estimates of the  critical exponents are  $\beta/\nu=0.477(7)$ and $\gamma/\nu=2.056(6)$~\cite{cunha2011critical}. The current estimates of the critical exponents of the three dimensional Ising model are $\nu=0.629971$, $\gamma/\nu=1.96370$, and $\beta/\nu=0.518149$~\cite{pelissetto2002critical}.

\begin{figure}
\includegraphics[width= \columnwidth]{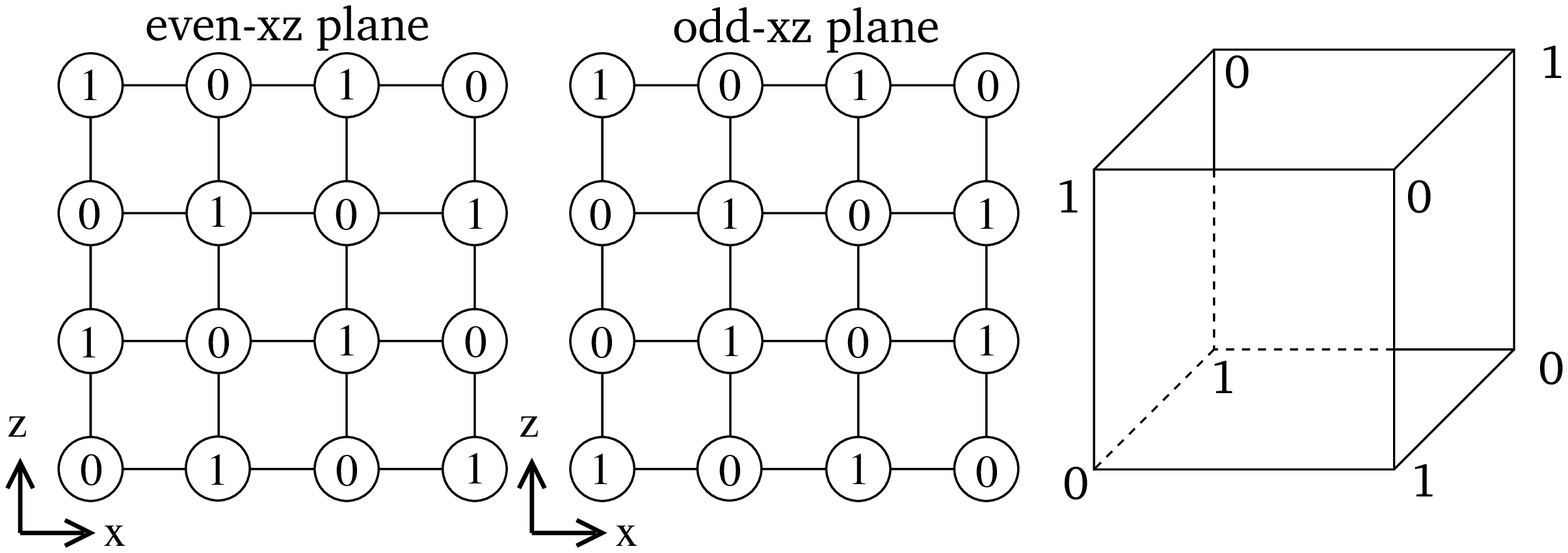}
\caption{For the $1$-NN model in three dimensions, the cubic lattice is  divided into two sublattices labeled by $0$ and $1$.}
\label{fig:3d1nn_sublattice}
\end{figure}

To define the order parameter, we divide the lattice into two sublattices  as shown in Fig.~\ref{fig:3d1nn_sublattice}. Each site of a certain sublattice is surrounded by six sites belonging to other sublattices. We define the order parameter $q_{1,3d}$  as
\be
q_{1,3d}=\left |\rho_0-\rho_1 \right |,
\ee
where $\rho_i$ denotes the densities of particles on sublattice $i$.  In the disordered phase $q_{1,3d}$ is zero, while in the sublattice phase $q_{1,3d}$ is nonzero.

\begin{figure}
\includegraphics[width= \columnwidth]{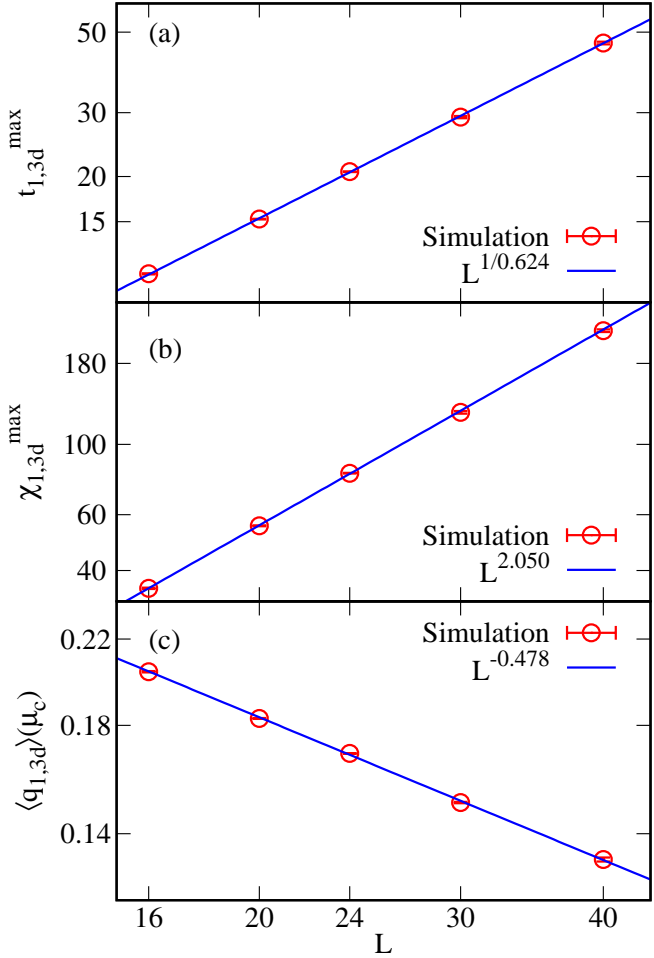}
\caption{(Color online) Power-law fits for the scaling of (a) $t_{1,3d}^{\mathrm{max}}$, (b) $\chi_{1,3d}^{\mathrm{max}}$, and (c) $\langle q_{1,3d} \rangle (\mu_c)$ with system size $L$  for the $1$-NN model in three dimensions. The axes are scaled logarithmically.} 
\label{fig:3d1nn_exponent}
\end{figure}

\begin{figure}
\includegraphics[width= \columnwidth]{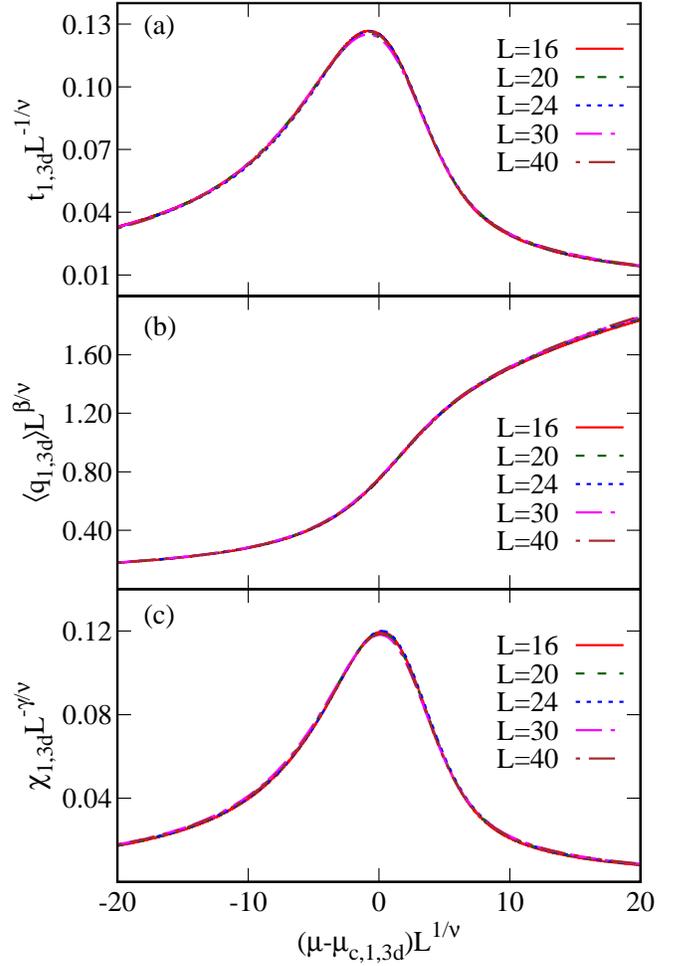}
\caption{ (Color online) Data for the $1$-NN model in three dimensions for different system sizes collapse onto one curve for (a) $t_{1,3d}$, (b) $\langle q_{1,3d} \rangle$, and (c) $\chi_{1,3d}$ when scaled as in Eq.~(\ref{eq:fullscaling}) with exponents $\nu = 0.624(5)$, $\beta/\nu=0.478(9)$, $\gamma/\nu=2.050(13)$, and $\mu_{c,1,3d} = 0.0558(6)$.}
\label{fig:3d1nn_collapse}
\end{figure}

We determine the density of states for system sizes up to $L=40$. 
We determine the critical exponents using Eq.~(\ref{eq:maxscaling}). The power-law scaling and the best fits are shown in Fig.~\ref{fig:3d1nn_exponent} for $t_{1,3d}^{\mathrm{max}}$,  $\chi_{1,3d}^{\mathrm{max}}$, and $q_{1,3d}(\mu_c(L))$. We obtain $\nu = 0.624(5)$, $\beta/\nu=0.478(9)$, and $\gamma/\nu=2.050(13)$.  Extrapolating $\mu_c(L)$ and $\eta_c(L)$ using Eqs.~(\ref{eq:muscaling}) and (\ref{eq:rhoscaling}), we obtain $\mu_{c,1,3d} = 0.0558(6)$ and $\rho_{c,1,3d}= 0.4220(2)$. These estimates are consistent with known estimates (see above) for the critical parameters.  The data for the thermodynamics quantities for different system sizes collapse onto one curve when scaled as in Eq.~(\ref{eq:fullscaling}) with the numerically obtained critical parameters (see Fig.~\ref{fig:3d1nn_collapse}).

\subsection{\label{sec:2nn3d}$2$-NN Model in three dimensions}
In the $2$-NN model in three dimensions, a particle excludes 18 sites from being occupied by another particle.  As density is increased, the  system undergoes a discontinuous phase transition from a low density disordered fluid phase to a high density ordered sublattice phase with bcc structure at full packing~\cite{2005-p-jcp-thermodynamic}. The estimates for the critical parameters are   $\mu_{c,2,3d}=0.53(1)$, with fluid and sublattice phases coexisting between $\rho_f=0.415(8)$ and  $\rho_s=0.515(8)$~\cite{2005-p-jcp-thermodynamic} (to convert from the notation in Ref.~\cite{2005-p-jcp-thermodynamic} to our notation,  $\beta \mu = \ln \sigma^3 + \mu_{c,2,3d}$ and $\sigma=\sqrt{3}$).

\begin{figure}
	\includegraphics[width=\columnwidth]{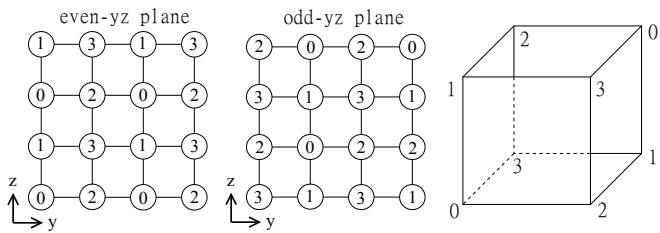}
	\caption{For the $2$-NN model in three dimensions, the cubic lattice is  divided into four sublattices labeled by $0$ to $3$. The diagonally opposite sites of each cube belong to the same sublattice.}
	\label{fig:3d2NNsublattice}
\end{figure}

To define the order parameter, we divide the lattice into four sublattices as shown in Fig.~\ref{fig:3d2NNsublattice}. The order parameter is defined as
\be
q_{2,3d}=\left| \sum_{j=0}^3 \rho_j \exp\left[j\frac{2\pi i}{4} \right] \right|,
\ee
where $\rho_j$ is the density of particles in sublattice $j$. When one of the sublattices is preferentially occupied, $q_{2,3d}$ becomes nonzero.

\begin{figure}
	\includegraphics[width=\columnwidth]{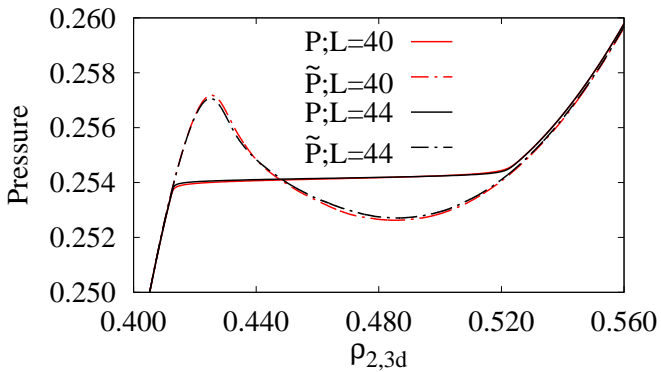}
	\caption{(Color online)  Variation of the grand canonical pressure $P$ computed from Eq.~(\ref{eq:P}) and the canonical pressure $\widetilde{P}$ computed from Eq.~(\ref{eq:Pphi}), with density for the $2$-NN model in three dimensions. The data are for the two largest system sizes studied.} 
	\label{fig:2nn3dP}
\end{figure}

We determine the density of states for system sizes up to $L = 44$.  We follow the same analysis as was done for the 3-NN model in two dimensions (see Sec.~\ref{sec:3nn2d}). The first order nature of the transition can be seen from studying pressure. Figure~\ref{fig:2nn3dP} shows the variation of pressure with density, computed both in the grand canonical ensemble ($P$) as well as the canonical ensemble ($\widetilde{P}$).  $\widetilde{P}$ is nonmonotonic, while $P$ is nearly a constant in the coexistence regime.  The curve for $P$ is similar to the usual Maxwell construction for a nonmonotonic $\widetilde{P}$.

 From the nonconvexity of the entropy, we estimated the coexistence densities $\rho_f(L)$ and $\rho_s(L)$ from the end points of the convex envelope and  critical chemical potential $\mu_c(L)$ using Eq.~(\ref{eqn:NC}). The critical parameters, thus obtained are tabulated in Table~\ref{tab:2nn3d}.
\begin{table}
	\caption{\label{tab:2nn3d} Critical parameters obtained from  nonconvexity of entropy  for the $2$-NN model in three dimensions. The data are extrapolated to infinite system size using  linear regression with $ L^{-3} $.} 
	\begin{ruledtabular}
		\begin{tabular}{c c c c}
		$L$& $\rho_f$ &$\rho_s$ &  $\mu_{c,2,2d}$  \\
		\hline
	%	16    & 0.1030273&0.1318359 &  0.529226   \\
	%	\hline
		20  &  0.41203(5) & 0.5238(1) & 0.53065(2) \\
		%\hline
		24  &  0.41243(3) & 0.5228(1) & 0.53150(2) \\
		%\hline
		28  &  0.41273(2) & 0.5222(1) & 0.53188(3)  \\
		%\hline
		32  & 0.41306(3) & 0.5212(1) & 0.53211(2)  \\
		%\hline
		36  &   0.41312(3) & 0.52068(6) & 0.53232(2)   \\
		%\hline
		40 & 0.41338(2) & 0.52050(6) & 0.53237(1) \\ %\hline
		44 & 0.41350(3) & 0.5203(1) & 0.53246(2) \\	
		%\hline \hline
		$ \infty $    &0.4136(1)& 0.5197(2)& 0.5326(4)\\  
	\end{tabular}
\end{ruledtabular}
\end{table}

\begin{figure}
	\includegraphics[width=\columnwidth]{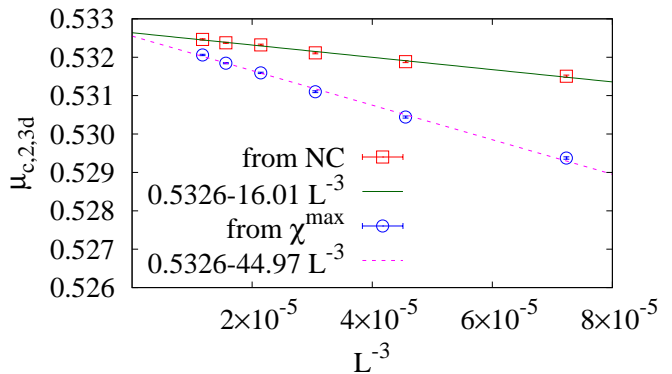}
	\caption{(Color online) Extrapolation of  critical chemical potential $\mu_{c,2,3d}(L)$ to infinite system size for the 2-NN model in three dimensions.}
	\label{fig:muc2nn3d}
\end{figure}

We extrapolate the critical parameters to infinite system size using Eqs.~(\ref{eq:muscaling}) and (\ref{eq:rhoscaling}) with $\nu=1/3$. The extrapolation for  $\mu_{c}(L)$ is shown in Fig.~\ref{fig:muc2nn3d}.
We obtain $\mu_{c,2,3d}=0.5326(4)$ from the nonconvex analysis and $\mu_{c,2,3d}=0.5326(3)$ from the analysis of susceptibility. Similarly, we obtain the coexistence densities in the thermodynamic limit to be $\rho_{f,2,3d}=0.4136(1)$ and  $\rho_{s,2,3d}=0.5197(2)$.  We also obtain the critical pressure to be $P_{c,2,3d} = 0.2542(1)$. These values should be compared with earlier estimates of $\mu_{c,2,3d}=0.53(1)$, $\rho_f=0.415(8)$, and  $\rho_s=0.515(8)$~\cite{2005-p-jcp-thermodynamic}. There is no earlier estimate of critical pressure.

\begin{figure}
	\includegraphics[width=\columnwidth]{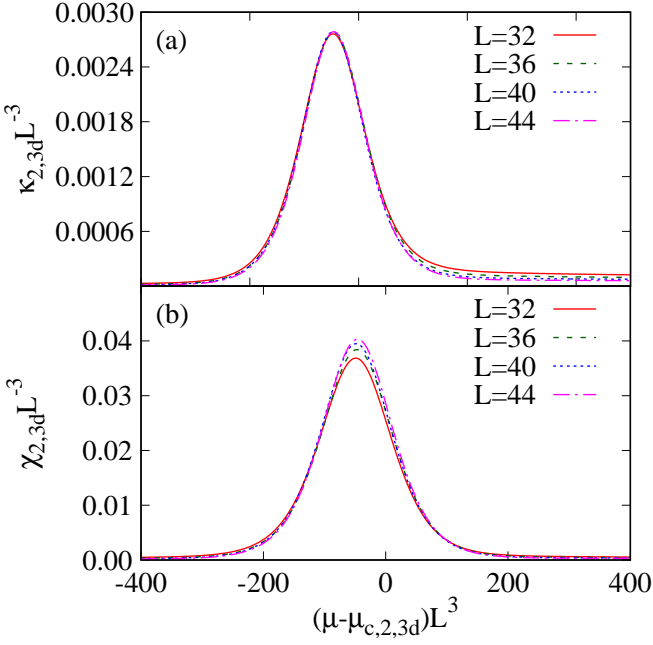}
	\caption{ (Color online) Data for different system sizes collapse for (a) $\kappa_{2,3d}$ and  (b) $\chi_{2,3d}$, when scaled as in Eq.~(\ref{eq:fullscaling}) with $\mu_{c,2,3d}=0.5326(4)$ and $\nu=1/d$ (for first order transition). The data are for the  $2$-NN model in three dimensions.}
	\label{fig:3d2NNcollapse}
\end{figure}

Finally, we show that the data for susceptibility and compressibility for different system sizes collapse onto one curve when scaled as in Eq.~(\ref{eq:fullscaling}) with the numerically obtained critical parameters and the exponents for a first order transition (see Fig.~\ref{fig:3d2NNcollapse}). We note that the data collapse for susceptibility has finite size corrections.

\subsection{\label{sec:3nn3d}$3$-NN Model in three dimensions}
 In the $3$-NN model in three dimensions, a particle excludes $26$ sites from being occupied by another particle. The model is equivalent to the model of $2\times2 \times 2$ hard cubes. The rich phase diagram of this model has been obtained recently based on extensive grand canonical Monte Carlo simulations with a transfer matrix based strip update algorithm~\cite{vigneshwar2019phase}.  The system undergoes three entropy driven phase transitions with increasing density: first from a disordered to a layered phase, second from the layered to a sublattice phase, and third from the sublattice to a columnar phase. Using finite-size scaling, it was shown that the disordered-layered phase transition is continuous, while the layered-sublattice and sublattice-columnar transitions are discontinuous~\cite{vigneshwar2019phase}.

To study the phase transitions, we define three order parameters $q^1_{3,3d}$, $q^2_{3,3d}$ and $q^3_{3,3d}$. We also define the density field $\eta(x,y,z)$ to  be $1$ if the site $(x,y,z)$ is occupied by a particle and $0$ otherwise. The Fourier transform of the density field $\tilde{\eta}(k_x, k_y,k_z)$ may be written as
\be
\tilde{\eta}(k_x, k_y,k_z) = \frac{8}{L^3}\,\sum_{x,y,z}\,\eta(x,y,z)\,\mathrm{e}^{i(k_x x + k_y y + k_z z)}
\label{eq:fourier}.
\ee
The vector order parameter $\mathbf L$ for the layered phase may be written as~\cite{vigneshwar2019phase}
\be
\mathbf{L} = (L_x, L_y, L_z),\label{eq:L}
\ee
where $L_x = \tilde{\eta}(\pi,0,0)$, $L_y = \tilde{\eta}(0,\pi,0)$ and $L_z=\tilde{\eta}(0,0,\pi)$.
Nonzero $L_x$, $L_y$, or $L_z$ implies that there is a translational order of period two in the $x$, $y$, or $z$ directions, respectively. The order parameters are then defined as
\bea
q^1_{3,3d}&=& \sqrt{L_x^2 + L_y^2 + L_z^2},\label{eq:q1}\\
q^2_{3,3d} &=& \sqrt{|\tilde{\eta}(\pi,\pi,0)|^2 + |\tilde{\eta}(0,\pi,\pi)|^2 + |\tilde{\eta}(\pi,0,\pi)|^2},\label{eq:q2}\\
q^3_{3,3d}&=& |\tilde{\eta}(\pi,\pi,\pi)|. \label{eq:q3}
\eea

\begin{figure}
\includegraphics[width= \columnwidth]{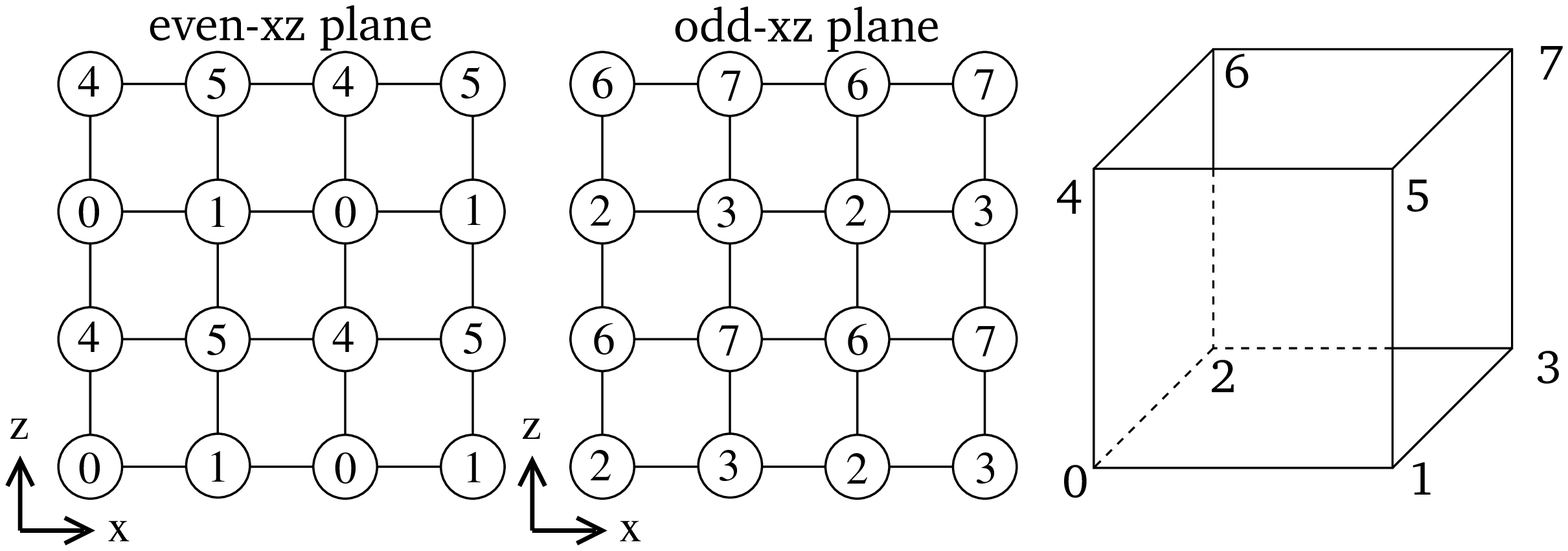}
\caption{ For the $3$-NN model in three dimensions, the cubic lattice is  divided into eight sublattices labeled by $0$ to $7$.}
\label{fig:3d3nn_sublattice}
\end{figure}

For a translationally invariant system, $q^1_{3,3d}$, $q^2_{3,3d}$, and  $q^3_{3,3d}$ are all zero. $q^1_{3,3d}$ is nonzero if there is a translational order in at least one of the three directions. $q^2_{3,3d}$ is nonzero if there is translational order in at least two of the three directions, while $q^3_{3,3d}$ is nonzero if there is translational order in all three directions. We divide the whole lattice into eight sublattices as shown in Fig.~\ref{fig:3d3nn_sublattice} and calculate the occupation densities of each type of sublattice. The order parameters defined in Eqs.~(\ref{eq:q1})--(\ref{eq:q3}) can be expressed in terms of eight sublattice densities.

\begin{figure}
  \includegraphics[width= \columnwidth]{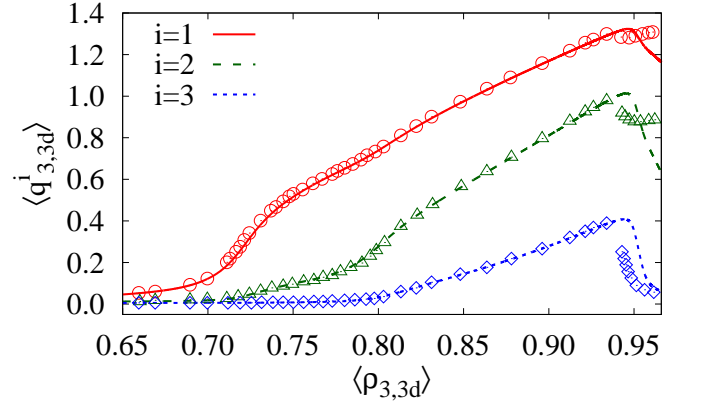}
\caption{(Color online) Comparison of the order parameters $q^1_{3,3d}$, $q^2_{3,3d}$, and $q^3_{3,3d}$ obtained from the flat histogram algorithm (lines) with those obtained from grand canonical Monte Carlo simulations~\cite{vigneshwar2019phase} (data points). The data are  for the $3$-NN model in three dimensions.}
\label{fig:q123}
\end{figure}

We find that it becomes difficult to flatten the histogram for this model, especially for larger system sizes. For this reason, for $L=50$, we stop after $17$ iterations.  
The variation of the three order parameters $q^1_{3,3d}$, $q^2_{3,3d}$, and $q^3_{3,3d}$ with $\rho$ is shown in Fig.~\ref{fig:q123} for system size $L=50$. The results are compared with results obtained from fixed chemical potential grand canonical simulations in Ref.~\cite{vigneshwar2019phase}. The data match very well for densities less than $0.92$. Beyond this density, all three order parameters show some discrepancy. In particular, we find that, in the flat histogram simulations, we obtain a layered phase at high densities while it should be columnar. The reason for this is that it is difficult to equilibrate the system at high densities. For example, in the grand canonical simulation, the equilibration time is order $10^7$ Monte Carlo steps~\cite{vigneshwar2019phase}. In the flat histogram algorithm, this is roughly the total time spent in an iteration, hence the difficulty with equilibration. However, we point out that the flat histogram algorithm is able to identify three phase transitions.
	
\section{\label{sec:summary}Summary and Conclusions}
In this paper, we implemented a flat histogram algorithm for hard-core lattice gases combining an efficient grand canonical transfer matrix based strip algorithm with the flat histogram Wang-Landau algorithm. We showed its efficacy by reproducing known results for the $k$-NN model for $k=1,2,3$ on the square and cubic lattices. These models covered a large number of scenarios: continuous phase transition, first order phase transitions, exponentially diverging entropy at full packing, and multiple phase transitions. Though the implementation is specific to these models, it can be easily generalized to hard-core lattice models of other shapes.

The implementation involves cluster moves that are rejection free. The current density of states are incorporated into the probabilities of choosing new configurations. This results in the low entropy state being accessed efficiently. In contrast, if a similar cluster move is applied but without biasing with the current density of states, then the algorithm fails to flatten the histogram, leading to significant errors. Thus the bias induced by including the density of states in the probability is crucial. Also, the implementation with the local single site evaporation-deposition moves fails to give results for larger $k$ or larger $L$, emphasizing the necessity of cluster moves. In addition, for the system sizes and values of $k$ for which all three algorithms give results, we showed that the error is minimum and the convergence is fastest for the strip update algorithm (SCWL).

We were able to estimate critical exponents of all continuous transitions with reasonable accuracy using SCWL. Also, for systems with large degeneracy in the ground state, SCWL is very efficient as shown for the $2$-NN model (Sec.~\ref{sec:2nn2d}). For the first order transitions in the 3-NN model in two dimensions (Sec.~\ref{sec:3nn2d}) and the 2-NN model in three dimensions  (Sec.~\ref{sec:2nn3d}) we could obtain improved estimates for the critical chemical potential and coexistence densities.  More recently, the SCWL algorithm has been used to obtain the detailed phase diagram of the lattice gas with third nearest neighbor exclusion on a triangular lattice~\cite{jaleel2021hard}.

While the flat histogram implementation was able to get accurate results for all the models studied, it may not be sufficient to obtain accurate results at high densities. For the model of hard cubes in three dimensions, which undergoes three phase transitions, the flat histogram appears to indicate a layered phase at densities close to full packing. However, the actual phase, obtained from fixed fugacity grand canonical simulations, has columnar nature. In the fixed fugacity simulations, at these densities, it takes an order of $10^{7}$ Monte Carlo steps to equilibrate the system. In the flat histogram implementation, during the random walk in configuration space, the system spends less time at a particular density. This is probably the reason for getting the phase wrong at high density for cubes. However, the flat histogram result does indicate a phase transition at the correct densities and one may have to supplement the result with fixed fugacity simulations to obtain more details.

 We showed that the entropy is nonconvex in the coexistence regime. The construction of the convex envelope gives excellent estimates for critical chemical potential as well as the coexistence densities for the $3$-NN model in two dimensions and the $2$-NN model in three dimensions.

A promising area for future study is binary gases. Here, exploring multidimensional phase space using fixed fugacity simulations is very time consuming. Flat histogram methods have a significant advantage in being able to access the full phase space in one sweep of the configuration space. The simplest model to study will be the mixture of $1$-NN and $0$-NN particles which shows a nontrivial phase diagram with a tricritical point~\cite{1983-p-jcp-coexistence,doi:10.1063/1.3658045,2015-os-pre-transfer,liu2001phase, rodrigues2019three, rodrigues2019thermodynamic, rodrigues2020fluid}. Estimating the critical parameters for the model from the flat histogram method would be a starting point.  It would also be interesting to implement the SCWL algorithm for spin systems with local interactions. Unlike the hard-core lattice gas system, this is a thermal system where the same methodology could be used in evaluating the density of states.

There are variants of the flat histogram method, for example, WL1/t, tomographic sampling, adaptive windows, etc. The implementation presented in this paper, which depends only on how the evaporation and deposition are implemented, will work for these variants also. Comparing the efficiency of the strip update algorithm for these flat histogram protocols would be interesting to study.

\begin{acknowledgments}
The simulations were carried out on the high performance computing machines Nandadevi at the Institute of Mathematical Sciences, Kalinga at National Institute of Science Education and Research (NISER), Bhubaneswar,  and the computational facilities provided by the University of Warwick Scientific Computing Research Technology Platform. J.E.T. thanks NISER for the hospitality during a visit when part of the work was done. 
\end{acknowledgments}

A.A.A.J. and J.E.T contributed equally to this work.

\appendix*

\section{\label{sec:appendix}DERIVATION of $C_{o}(\ell,n)$ AND $C_{p}(\ell,n)$}

In this appendix, we outline the derivation of $C_{o}(\ell,n)$ and $C_{p}(\ell,n)$, the number of ways of filling a one dimensional lattice of $\ell$ sites with $n$ particles with nearest neighbor exclusion with open and periodic boundary condition, respectively.
Consider first $C_{o}(\ell,n)$. The set of configurations can be broken into configurations where the last site is empty [denote these by $E_{o}(\ell,n)$] and those configurations where the last site is filled. The latter corresponds to configurations where the last but one site is empty. Thus
\be
C_{o}(\ell,n) = E_{o}(\ell,n)+E_{o}(\ell-1,n-1).
\label{eqn:C_{o}_sum} 
\ee
The enumeration of $E_{o}(\ell,n)$ is equivalent to the arrangement of $n$ dimers and $\ell-2n$ holes, and hence
\be
E_{o}(\ell,n) =  \frac{(\ell-n)!}{(\ell-2n)!n!}.
\label{eqn:C_{o}_term}
\ee
Using Eqs.~(\ref{eqn:C_{o}_sum}) and~(\ref{eqn:C_{o}_term}), we immediately obtain 
\be
C_{o}(\ell,n) = \frac{(\ell-n+1)!}{(\ell-2n+1)!n!}.	
\label{eqn:C_{o}_final}
\ee
Now consider $C_{p}(\ell,n)$ for a periodic ring. Choose a site at random. This site could be either filled or empty, both cases reducing to the problem of a segment with open boundary conditions:
\be
C_{p}(\ell,n) = C_{o}(\ell-1,n)+C_{o}(\ell-3,n-1). 
\label{eqn:C_{p}_sum}
\ee
Using Eqs.~(\ref{eqn:C_{o}_final}) and~(\ref{eqn:C_{p}_sum}), we obtain
\be
C_{p}(\ell,n) =  \frac{\ell(\ell-n-1)!}{(\ell-2n)!n!}.
\label{eqn:C_{p}_final}
\ee


\begin{thebibliography}{83}%
	\makeatletter
	\providecommand \@ifxundefined [1]{%
		\@ifx{#1\undefined}
	}%
	\providecommand \@ifnum [1]{%
		\ifnum #1\expandafter \@firstoftwo
		\else \expandafter \@secondoftwo
		\fi
	}%
	\providecommand \@ifx [1]{%
		\ifx #1\expandafter \@firstoftwo
		\else \expandafter \@secondoftwo
		\fi
	}%
	\providecommand \natexlab [1]{#1}%
	\providecommand \enquote  [1]{``#1''}%
	\providecommand \bibnamefont  [1]{#1}%
	\providecommand \bibfnamefont [1]{#1}%
	\providecommand \citenamefont [1]{#1}%
	\providecommand \href@noop [0]{\@secondoftwo}%
	\providecommand \href [0]{\begingroup \@sanitize@url \@href}%
	\providecommand \@href[1]{\@@startlink{#1}\@@href}%
	\providecommand \@@href[1]{\endgroup#1\@@endlink}%
	\providecommand \@sanitize@url [0]{\catcode `\\12\catcode `\$12\catcode
		`\&12\catcode `\#12\catcode `\^12\catcode `\_12\catcode `\%12\relax}%
	\providecommand \@@startlink[1]{}%
	\providecommand \@@endlink[0]{}%
	\providecommand \url  [0]{\begingroup\@sanitize@url \@url }%
	\providecommand \@url [1]{\endgroup\@href {#1}{\urlprefix }}%
	\providecommand \urlprefix  [0]{URL }%
	\providecommand \Eprint [0]{\href }%
	\providecommand \doibase [0]{https://doi.org/}%
	\providecommand \selectlanguage [0]{\@gobble}%
	\providecommand \bibinfo  [0]{\@secondoftwo}%
	\providecommand \bibfield  [0]{\@secondoftwo}%
	\providecommand \translation [1]{[#1]}%
	\providecommand \BibitemOpen [0]{}%
	\providecommand \bibitemStop [0]{}%
	\providecommand \bibitemNoStop [0]{.\EOS\space}%
	\providecommand \EOS [0]{\spacefactor3000\relax}%
	\providecommand \BibitemShut  [1]{\csname bibitem#1\endcsname}%
	\let\auto@bib@innerbib\@empty
	%</preamble>
	\bibitem [{\citenamefont {Runnels}(1972)}]{runnels1972phase}%
	\BibitemOpen
	\bibfield  {author} {\bibinfo {author} {\bibfnamefont {L.}~\bibnamefont
			{Runnels}},\ }\bibfield  {title} {\bibinfo {title} {Phase transitions and
			critical phenomena},\ }\href@noop {} {\bibfield  {journal} {\bibinfo
			{journal} {Domb M. S. Green Eds}\ }\textbf {\bibinfo {volume} {2}},\ \bibinfo
		{pages} {305} (\bibinfo {year} {1972})}\BibitemShut {NoStop}%
	\bibitem [{\citenamefont {Alder}\ and\ \citenamefont
		{Wainwright}(1957)}]{1957-aw-jcp-phase}%
	\BibitemOpen
	\bibfield  {author} {\bibinfo {author} {\bibfnamefont {B.~J.}\ \bibnamefont
			{Alder}}\ and\ \bibinfo {author} {\bibfnamefont {T.~E.}\ \bibnamefont
			{Wainwright}},\ }\bibfield  {title} {\bibinfo {title} {Phase transition for a
			hard sphere system},\ }\href {https://doi.org/10.1063/1.1743957} {\bibfield
		{journal} {\bibinfo  {journal} {J. Chem. Phys.}\ }\textbf {\bibinfo {volume}
			{27}},\ \bibinfo {pages} {1208} (\bibinfo {year} {1957})}\BibitemShut
	{NoStop}%
	\bibitem [{\citenamefont {Alder}\ and\ \citenamefont
		{Wainwright}(1962)}]{1962-aw-pr-phase}%
	\BibitemOpen
	\bibfield  {author} {\bibinfo {author} {\bibfnamefont {B.~J.}\ \bibnamefont
			{Alder}}\ and\ \bibinfo {author} {\bibfnamefont {T.~E.}\ \bibnamefont
			{Wainwright}},\ }\bibfield  {title} {\bibinfo {title} {Phase transition in
			elastic disks},\ }\href {https://doi.org/10.1103/PhysRev.127.359} {\bibfield
		{journal} {\bibinfo  {journal} {Phys. Rev.}\ }\textbf {\bibinfo {volume}
			{127}},\ \bibinfo {pages} {359} (\bibinfo {year} {1962})}\BibitemShut
	{NoStop}%
	\bibitem [{\citenamefont {Cuetos}\ \emph {et~al.}(2017)\citenamefont {Cuetos},
		\citenamefont {Dennison}, \citenamefont {Masters},\ and\ \citenamefont
		{Patti}}]{cuetos-2007-phase}%
	\BibitemOpen
	\bibfield  {author} {\bibinfo {author} {\bibfnamefont {A.}~\bibnamefont
			{Cuetos}}, \bibinfo {author} {\bibfnamefont {M.}~\bibnamefont {Dennison}},
		\bibinfo {author} {\bibfnamefont {A.}~\bibnamefont {Masters}},\ and\ \bibinfo
		{author} {\bibfnamefont {A.}~\bibnamefont {Patti}},\ }\bibfield  {title}
	{\bibinfo {title} {Phase behaviour of hard board-like particles},\ }\href
	{https://doi.org/10.1039/C7SM00726D} {\bibfield  {journal} {\bibinfo
			{journal} {Soft Matter}\ }\textbf {\bibinfo {volume} {13}},\ \bibinfo {pages}
		{4720} (\bibinfo {year} {2017})}\BibitemShut {NoStop}%
	\bibitem [{\citenamefont {Mirzad~Rafael}\ \emph {et~al.}(2020)\citenamefont
		{Mirzad~Rafael}, \citenamefont {Corbett}, \citenamefont {Cuetos},\ and\
		\citenamefont {Patti}}]{rafael-2020-self}%
	\BibitemOpen
	\bibfield  {author} {\bibinfo {author} {\bibfnamefont {E.}~\bibnamefont
			{Mirzad~Rafael}}, \bibinfo {author} {\bibfnamefont {D.}~\bibnamefont
			{Corbett}}, \bibinfo {author} {\bibfnamefont {A.}~\bibnamefont {Cuetos}},\
		and\ \bibinfo {author} {\bibfnamefont {A.}~\bibnamefont {Patti}},\ }\bibfield
	{title} {\bibinfo {title} {Self-assembly of freely-rotating polydisperse
			cuboids: unveiling the boundaries of the biaxial nematic phase},\ }\href
	{https://doi.org/10.1039/D0SM00484G} {\bibfield  {journal} {\bibinfo
			{journal} {Soft Matter}\ }\textbf {\bibinfo {volume} {16}},\ \bibinfo {pages}
		{5565} (\bibinfo {year} {2020})}\BibitemShut {NoStop}%
	\bibitem [{\citenamefont {Taylor}\ \emph {et~al.}(1985)\citenamefont {Taylor},
		\citenamefont {Williams}, \citenamefont {Park}, \citenamefont {Bartelt},\
		and\ \citenamefont {Einstein}}]{1985-prb-twpbe-two}%
	\BibitemOpen
	\bibfield  {author} {\bibinfo {author} {\bibfnamefont {D.~E.}\ \bibnamefont
			{Taylor}}, \bibinfo {author} {\bibfnamefont {E.~D.}\ \bibnamefont
			{Williams}}, \bibinfo {author} {\bibfnamefont {R.~L.}\ \bibnamefont {Park}},
		\bibinfo {author} {\bibfnamefont {N.~C.}\ \bibnamefont {Bartelt}},\ and\
		\bibinfo {author} {\bibfnamefont {T.~L.}\ \bibnamefont {Einstein}},\
	}\bibfield  {title} {\bibinfo {title} {Two-dimensional ordering of chlorine
			on ag(100)},\ }\href {https://doi.org/10.1103/PhysRevB.32.4653} {\bibfield
		{journal} {\bibinfo  {journal} {Phys. Rev. B}\ }\textbf {\bibinfo {volume}
			{32}},\ \bibinfo {pages} {4653} (\bibinfo {year} {1985})}\BibitemShut
	{NoStop}%
	\bibitem [{\citenamefont {Bak}\ \emph {et~al.}(1985)\citenamefont {Bak},
		\citenamefont {Kleban}, \citenamefont {Unertl}, \citenamefont {Ochab},
		\citenamefont {Akinci}, \citenamefont {Bartelt},\ and\ \citenamefont
		{Einstein}}]{1985-bkuoabe-prl-phase}%
	\BibitemOpen
	\bibfield  {author} {\bibinfo {author} {\bibfnamefont {P.}~\bibnamefont
			{Bak}}, \bibinfo {author} {\bibfnamefont {P.}~\bibnamefont {Kleban}},
		\bibinfo {author} {\bibfnamefont {W.~N.}\ \bibnamefont {Unertl}}, \bibinfo
		{author} {\bibfnamefont {J.}~\bibnamefont {Ochab}}, \bibinfo {author}
		{\bibfnamefont {G.}~\bibnamefont {Akinci}}, \bibinfo {author} {\bibfnamefont
			{N.~C.}\ \bibnamefont {Bartelt}},\ and\ \bibinfo {author} {\bibfnamefont
			{T.~L.}\ \bibnamefont {Einstein}},\ }\bibfield  {title} {\bibinfo {title}
		{Phase diagram of selenium adsorbed on the ni(100) surface: A physical
			realization of the ashkin-teller model},\ }\href
	{https://doi.org/10.1103/PhysRevLett.54.1539} {\bibfield  {journal} {\bibinfo
			{journal} {Phys. Rev. Lett.}\ }\textbf {\bibinfo {volume} {54}},\ \bibinfo
		{pages} {1539} (\bibinfo {year} {1985})}\BibitemShut {NoStop}%
	\bibitem [{\citenamefont {Dhar}(1982)}]{1982-d-prl-equivalence}%
	\BibitemOpen
	\bibfield  {author} {\bibinfo {author} {\bibfnamefont {D.}~\bibnamefont
			{Dhar}},\ }\bibfield  {title} {\bibinfo {title} {Equivalence of the
			two-dimensional directed-site animal problem to baxter's hard-square
			lattice-gas model},\ }\href {https://doi.org/10.1103/PhysRevLett.49.959}
	{\bibfield  {journal} {\bibinfo  {journal} {Phys. Rev. Lett.}\ }\textbf
		{\bibinfo {volume} {49}},\ \bibinfo {pages} {959} (\bibinfo {year}
		{1982})}\BibitemShut {NoStop}%
	\bibitem [{\citenamefont {Dhar}(1983)}]{1983-d-prl-exact}%
	\BibitemOpen
	\bibfield  {author} {\bibinfo {author} {\bibfnamefont {D.}~\bibnamefont
			{Dhar}},\ }\bibfield  {title} {\bibinfo {title} {Exact solution of a
			directed-site animals-enumeration problem in three dimensions},\ }\href
	{https://doi.org/10.1103/PhysRevLett.51.853} {\bibfield  {journal} {\bibinfo
			{journal} {Phys. Rev. Lett.}\ }\textbf {\bibinfo {volume} {51}},\ \bibinfo
		{pages} {853} (\bibinfo {year} {1983})}\BibitemShut {NoStop}%
	\bibitem [{\citenamefont {Brydges}\ and\ \citenamefont
		{Imbrie}(2003)}]{2003-bi-jsp-dimensional}%
	\BibitemOpen
	\bibfield  {author} {\bibinfo {author} {\bibfnamefont {D.~C.}\ \bibnamefont
			{Brydges}}\ and\ \bibinfo {author} {\bibfnamefont {J.~Z.}\ \bibnamefont
			{Imbrie}},\ }\bibfield  {title} {\bibinfo {title} {Dimensional reduction
			formulas for branched polymer correlation functions},\ }\href
	{https://doi.org/10.1023/A:1022143331697} {\bibfield  {journal} {\bibinfo
			{journal} {J. Stat. Phys.}\ }\textbf {\bibinfo {volume} {110}},\ \bibinfo
		{pages} {503} (\bibinfo {year} {2003})}\BibitemShut {NoStop}%
	\bibitem [{\citenamefont {Parisi}\ and\ \citenamefont
		{Sourlas}(1981)}]{1981-ps-prl-critical}%
	\BibitemOpen
	\bibfield  {author} {\bibinfo {author} {\bibfnamefont {G.}~\bibnamefont
			{Parisi}}\ and\ \bibinfo {author} {\bibfnamefont {N.}~\bibnamefont
			{Sourlas}},\ }\bibfield  {title} {\bibinfo {title} {Critical behavior of
			branched polymers and the lee-yang edge singularity},\ }\href
	{https://doi.org/10.1103/PhysRevLett.46.871} {\bibfield  {journal} {\bibinfo
			{journal} {Phys. Rev. Lett.}\ }\textbf {\bibinfo {volume} {46}},\ \bibinfo
		{pages} {871} (\bibinfo {year} {1981})}\BibitemShut {NoStop}%
	\bibitem [{\citenamefont {Flory}(1956)}]{1956-f-prs-phase}%
	\BibitemOpen
	\bibfield  {author} {\bibinfo {author} {\bibfnamefont {P.~J.}\ \bibnamefont
			{Flory}},\ }\bibfield  {title} {\bibinfo {title} {Phase equilibria in
			solutions of rod-like particles},\ }\href
	{https://doi.org/10.1098/rspa.1956.0016} {\bibfield  {journal} {\bibinfo
			{journal} {Proc. Roy. Soc. A}\ }\textbf {\bibinfo {volume} {234}},\ \bibinfo
		{pages} {73} (\bibinfo {year} {1956})}\BibitemShut {NoStop}%
	\bibitem [{\citenamefont {Ghosh}\ and\ \citenamefont
		{Dhar}(2007)}]{2007-gd-epl-on}%
	\BibitemOpen
	\bibfield  {author} {\bibinfo {author} {\bibfnamefont {A.}~\bibnamefont
			{Ghosh}}\ and\ \bibinfo {author} {\bibfnamefont {D.}~\bibnamefont {Dhar}},\
	}\bibfield  {title} {\bibinfo {title} {On the orientational ordering of long
			rods on a lattice},\ }\href {http://stacks.iop.org/0295-5075/78/i=2/a=20003}
	{\bibfield  {journal} {\bibinfo  {journal} {Eur. Phys. Lett.}\ }\textbf
		{\bibinfo {volume} {78}},\ \bibinfo {pages} {20003} (\bibinfo {year}
		{2007})}\BibitemShut {NoStop}%
	\bibitem [{\citenamefont {Kundu}\ \emph {et~al.}(2013)\citenamefont {Kundu},
		\citenamefont {Rajesh}, \citenamefont {Dhar},\ and\ \citenamefont
		{Stilck}}]{2013-krds-pre-nematic}%
	\BibitemOpen
	\bibfield  {author} {\bibinfo {author} {\bibfnamefont {J.}~\bibnamefont
			{Kundu}}, \bibinfo {author} {\bibfnamefont {R.}~\bibnamefont {Rajesh}},
		\bibinfo {author} {\bibfnamefont {D.}~\bibnamefont {Dhar}},\ and\ \bibinfo
		{author} {\bibfnamefont {J.~F.}\ \bibnamefont {Stilck}},\ }\bibfield  {title}
	{\bibinfo {title} {Nematic-disordered phase transition in systems of long
			rigid rods on two-dimensional lattices},\ }\href
	{https://doi.org/10.1103/PhysRevE.87.032103} {\bibfield  {journal} {\bibinfo
			{journal} {Phys. Rev. E}\ }\textbf {\bibinfo {volume} {87}},\ \bibinfo
		{pages} {032103} (\bibinfo {year} {2013})}\BibitemShut {NoStop}%
	\bibitem [{\citenamefont {Gschwind}\ \emph {et~al.}(2017)\citenamefont
		{Gschwind}, \citenamefont {Klopotek}, \citenamefont {Ai},\ and\ \citenamefont
		{Oettel}}]{2017-gkao-pre-isotropic}%
	\BibitemOpen
	\bibfield  {author} {\bibinfo {author} {\bibfnamefont {A.}~\bibnamefont
			{Gschwind}}, \bibinfo {author} {\bibfnamefont {M.}~\bibnamefont {Klopotek}},
		\bibinfo {author} {\bibfnamefont {Y.}~\bibnamefont {Ai}},\ and\ \bibinfo
		{author} {\bibfnamefont {M.}~\bibnamefont {Oettel}},\ }\bibfield  {title}
	{\bibinfo {title} {Isotropic-nematic transition for hard rods on a
			three-dimensional cubic lattice},\ }\href
	{https://doi.org/10.1103/PhysRevE.96.012104} {\bibfield  {journal} {\bibinfo
			{journal} {Phys. Rev. E}\ }\textbf {\bibinfo {volume} {96}},\ \bibinfo
		{pages} {012104} (\bibinfo {year} {2017})}\BibitemShut {NoStop}%
	\bibitem [{\citenamefont {Vigneshwar}\ \emph {et~al.}(2017)\citenamefont
		{Vigneshwar}, \citenamefont {Dhar},\ and\ \citenamefont
		{Rajesh}}]{2017-vdr-jsm-different}%
	\BibitemOpen
	\bibfield  {author} {\bibinfo {author} {\bibfnamefont {N.}~\bibnamefont
			{Vigneshwar}}, \bibinfo {author} {\bibfnamefont {D.}~\bibnamefont {Dhar}},\
		and\ \bibinfo {author} {\bibfnamefont {R.}~\bibnamefont {Rajesh}},\
	}\bibfield  {title} {\bibinfo {title} {Different phases of a system of hard
			rods on three dimensional cubic lattice},\ }\href
	{http://stacks.iop.org/1742-5468/2017/i=11/a=113304} {\bibfield  {journal}
		{\bibinfo  {journal} {J. Stat. Mech.}\ }\textbf {\bibinfo {volume} {2017}},\
		\bibinfo {pages} {113304} (\bibinfo {year} {2017})}\BibitemShut {NoStop}%
	\bibitem [{\citenamefont {Mao}\ \emph {et~al.}(2002)\citenamefont {Mao},
		\citenamefont {Harris},\ and\ \citenamefont {Stine}}]{2002-mhs-jcp-simple}%
	\BibitemOpen
	\bibfield  {author} {\bibinfo {author} {\bibfnamefont {L.}~\bibnamefont
			{Mao}}, \bibinfo {author} {\bibfnamefont {H.~H.}\ \bibnamefont {Harris}},\
		and\ \bibinfo {author} {\bibfnamefont {K.~J.}\ \bibnamefont {Stine}},\
	}\bibfield  {title} {\bibinfo {title} {Simple lattice simulation of chiral
			discrimination in monolayers},\ }\href@noop {} {\bibfield  {journal}
		{\bibinfo  {journal} {J. Chem. Inform. Comput. Sci.}\ }\textbf {\bibinfo
			{volume} {42}},\ \bibinfo {pages} {1179} (\bibinfo {year}
		{2002})}\BibitemShut {NoStop}%
	\bibitem [{\citenamefont {Barnes}\ \emph {et~al.}(2009)\citenamefont {Barnes},
		\citenamefont {Siderius},\ and\ \citenamefont
		{Gelb}}]{2009-bsg-langmuir-structure}%
	\BibitemOpen
	\bibfield  {author} {\bibinfo {author} {\bibfnamefont {B.~C.}\ \bibnamefont
			{Barnes}}, \bibinfo {author} {\bibfnamefont {D.~W.}\ \bibnamefont
			{Siderius}},\ and\ \bibinfo {author} {\bibfnamefont {L.~D.}\ \bibnamefont
			{Gelb}},\ }\bibfield  {title} {\bibinfo {title} {Structure, thermodynamics,
			and solubility in tetromino fluids},\ }\href
	{https://doi.org/10.1021/la900196b} {\bibfield  {journal} {\bibinfo
			{journal} {Langmuir}\ }\textbf {\bibinfo {volume} {25}},\ \bibinfo {pages}
		{6702} (\bibinfo {year} {2009})}\BibitemShut {NoStop}%
	\bibitem [{\citenamefont {Verberkmoes}\ and\ \citenamefont
		{Nienhuis}(1999)}]{1999-vn-prl-triangular}%
	\BibitemOpen
	\bibfield  {author} {\bibinfo {author} {\bibfnamefont {A.}~\bibnamefont
			{Verberkmoes}}\ and\ \bibinfo {author} {\bibfnamefont {B.}~\bibnamefont
			{Nienhuis}},\ }\bibfield  {title} {\bibinfo {title} {Triangular trimers on
			the triangular lattice: An exact solution},\ }\href
	{https://doi.org/10.1103/PhysRevLett.83.3986} {\bibfield  {journal} {\bibinfo
			{journal} {Phys. Rev. Lett.}\ }\textbf {\bibinfo {volume} {83}},\ \bibinfo
		{pages} {3986} (\bibinfo {year} {1999})}\BibitemShut {NoStop}%
	\bibitem [{\citenamefont {Szabelski}\ \emph {et~al.}(2013)\citenamefont
		{Szabelski}, \citenamefont {Rzysko}, \citenamefont {Panczyk}, \citenamefont
		{Ghijsens}, \citenamefont {Tahara}, \citenamefont {Tobe},\ and\ \citenamefont
		{De~Feyter}}]{szabelski2013selfassembly}%
	\BibitemOpen
	\bibfield  {author} {\bibinfo {author} {\bibfnamefont {P.}~\bibnamefont
			{Szabelski}}, \bibinfo {author} {\bibfnamefont {W.}~\bibnamefont {Rzysko}},
		\bibinfo {author} {\bibfnamefont {T.}~\bibnamefont {Panczyk}}, \bibinfo
		{author} {\bibfnamefont {E.}~\bibnamefont {Ghijsens}}, \bibinfo {author}
		{\bibfnamefont {K.}~\bibnamefont {Tahara}}, \bibinfo {author} {\bibfnamefont
			{Y.}~\bibnamefont {Tobe}},\ and\ \bibinfo {author} {\bibfnamefont
			{S.}~\bibnamefont {De~Feyter}},\ }\bibfield  {title} {\bibinfo {title}
		{Self-assembly of molecular tripods in two dimensions: structure and
			thermodynamics from computer simulations},\ }\href@noop {} {\bibfield
		{journal} {\bibinfo  {journal} {RSC Adv.}\ }\textbf {\bibinfo {volume} {3}},\
		\bibinfo {pages} {25159} (\bibinfo {year} {2013})}\BibitemShut {NoStop}%
	\bibitem [{\citenamefont {Ruth}\ \emph {et~al.}(2015)\citenamefont {Ruth},
		\citenamefont {Toral}, \citenamefont {Holz}, \citenamefont {Rickman},\ and\
		\citenamefont {Gunton}}]{2015-rthrg-tsf-impact}%
	\BibitemOpen
	\bibfield  {author} {\bibinfo {author} {\bibfnamefont {D.}~\bibnamefont
			{Ruth}}, \bibinfo {author} {\bibfnamefont {R.}~\bibnamefont {Toral}},
		\bibinfo {author} {\bibfnamefont {D.}~\bibnamefont {Holz}}, \bibinfo {author}
		{\bibfnamefont {J.}~\bibnamefont {Rickman}},\ and\ \bibinfo {author}
		{\bibfnamefont {J.}~\bibnamefont {Gunton}},\ }\bibfield  {title} {\bibinfo
		{title} {Impact of surface interactions on the phase behavior of y-shaped
			molecules},\ }\href
	{https://doi.org/https://doi.org/10.1016/j.tsf.2015.11.046} {\bibfield
		{journal} {\bibinfo  {journal} {Thin Solid Films}\ }\textbf {\bibinfo
			{volume} {597}},\ \bibinfo {pages} {188 } (\bibinfo {year}
		{2015})}\BibitemShut {NoStop}%
	\bibitem [{\citenamefont {Mandal}\ \emph {et~al.}(2018)\citenamefont {Mandal},
		\citenamefont {Nath},\ and\ \citenamefont {Rajesh}}]{2018-pre-mnr-phase}%
	\BibitemOpen
	\bibfield  {author} {\bibinfo {author} {\bibfnamefont {D.}~\bibnamefont
			{Mandal}}, \bibinfo {author} {\bibfnamefont {T.}~\bibnamefont {Nath}},\ and\
		\bibinfo {author} {\bibfnamefont {R.}~\bibnamefont {Rajesh}},\ }\bibfield
	{title} {\bibinfo {title} {Phase transitions in a system of hard y-shaped
			particles on the triangular lattice},\ }\href
	{https://doi.org/10.1103/PhysRevE.97.032131} {\bibfield  {journal} {\bibinfo
			{journal} {Phys. Rev. E}\ }\textbf {\bibinfo {volume} {97}},\ \bibinfo
		{pages} {032131} (\bibinfo {year} {2018})}\BibitemShut {NoStop}%
	\bibitem [{\citenamefont {Baxter}(1980)}]{1980-b-jpa-hard}%
	\BibitemOpen
	\bibfield  {author} {\bibinfo {author} {\bibfnamefont {R.~J.}\ \bibnamefont
			{Baxter}},\ }\bibfield  {title} {\bibinfo {title} {Hard hexagons: exact
			solution},\ }\href {http://stacks.iop.org/0305-4470/13/i=3/a=007} {\bibfield
		{journal} {\bibinfo  {journal} {J. Phys. A}\ }\textbf {\bibinfo {volume}
			{13}},\ \bibinfo {pages} {L61} (\bibinfo {year} {1980})}\BibitemShut
	{NoStop}%
	\bibitem [{\citenamefont {Vigneshwar}\ \emph {et~al.}(2019)\citenamefont
		{Vigneshwar}, \citenamefont {Mandal}, \citenamefont {Damle}, \citenamefont
		{Dhar},\ and\ \citenamefont {Rajesh}}]{vigneshwar2019phase}%
	\BibitemOpen
	\bibfield  {author} {\bibinfo {author} {\bibfnamefont {N.}~\bibnamefont
			{Vigneshwar}}, \bibinfo {author} {\bibfnamefont {D.}~\bibnamefont {Mandal}},
		\bibinfo {author} {\bibfnamefont {K.}~\bibnamefont {Damle}}, \bibinfo
		{author} {\bibfnamefont {D.}~\bibnamefont {Dhar}},\ and\ \bibinfo {author}
		{\bibfnamefont {R.}~\bibnamefont {Rajesh}},\ }\bibfield  {title} {\bibinfo
		{title} {Phase diagram of a system of hard cubes on the cubic lattice},\
	}\href@noop {} {\bibfield  {journal} {\bibinfo  {journal} {Phys. Rev. E}\
		}\textbf {\bibinfo {volume} {99}},\ \bibinfo {pages} {052129} (\bibinfo
		{year} {2019})}\BibitemShut {NoStop}%
	\bibitem [{\citenamefont {Kundu}\ and\ \citenamefont
		{Rajesh}(2014)}]{2014-kr-pre-phase}%
	\BibitemOpen
	\bibfield  {author} {\bibinfo {author} {\bibfnamefont {J.}~\bibnamefont
			{Kundu}}\ and\ \bibinfo {author} {\bibfnamefont {R.}~\bibnamefont {Rajesh}},\
	}\bibfield  {title} {\bibinfo {title} {Phase transitions in a system of hard
			rectangles on the square lattice},\ }\href
	{https://doi.org/10.1103/PhysRevE.89.052124} {\bibfield  {journal} {\bibinfo
			{journal} {Phys. Rev. E}\ }\textbf {\bibinfo {volume} {89}},\ \bibinfo
		{pages} {052124} (\bibinfo {year} {2014})}\BibitemShut {NoStop}%
	\bibitem [{\citenamefont {Kundu}\ and\ \citenamefont
		{Rajesh}(2015)}]{2015-kr-epjb-phase}%
	\BibitemOpen
	\bibfield  {author} {\bibinfo {author} {\bibfnamefont {J.}~\bibnamefont
			{Kundu}}\ and\ \bibinfo {author} {\bibfnamefont {R.}~\bibnamefont {Rajesh}},\
	}\bibfield  {title} {\bibinfo {title} {Phase transitions in systems of hard
			rectangles with non-integer aspect ratio},\ }\href
	{https://doi.org/10.1140/epjb/e2015-60210-7} {\bibfield  {journal} {\bibinfo
			{journal} {Eur. Phys. J. B}\ }\textbf {\bibinfo {volume} {88}},\ \bibinfo
		{pages} {133} (\bibinfo {year} {2015})}\BibitemShut {NoStop}%
	\bibitem [{\citenamefont {Fernandes}\ \emph {et~al.}(2007)\citenamefont
		{Fernandes}, \citenamefont {Arenzon},\ and\ \citenamefont
		{Levin}}]{2007-fal-jcp-monte}%
	\BibitemOpen
	\bibfield  {author} {\bibinfo {author} {\bibfnamefont {H.~C.~M.}\
			\bibnamefont {Fernandes}}, \bibinfo {author} {\bibfnamefont {J.~J.}\
			\bibnamefont {Arenzon}},\ and\ \bibinfo {author} {\bibfnamefont
			{Y.}~\bibnamefont {Levin}},\ }\bibfield  {title} {\bibinfo {title} {Monte
			carlo simulations of two-dimensional hard core lattice gases},\ }\href
	{https://doi.org/10.1063/1.2539141} {\bibfield  {journal} {\bibinfo
			{journal} {J. Chem. Phys.}\ }\textbf {\bibinfo {volume} {126}},\ \bibinfo
		{pages} {114508} (\bibinfo {year} {2007})}\BibitemShut {NoStop}%
	\bibitem [{\citenamefont {Nath}\ and\ \citenamefont
		{Rajesh}(2014)}]{2014-nr-pre-multiple}%
	\BibitemOpen
	\bibfield  {author} {\bibinfo {author} {\bibfnamefont {T.}~\bibnamefont
			{Nath}}\ and\ \bibinfo {author} {\bibfnamefont {R.}~\bibnamefont {Rajesh}},\
	}\bibfield  {title} {\bibinfo {title} {Multiple phase transitions in extended
			hard-core lattice gas models in two dimensions},\ }\href
	{https://doi.org/10.1103/PhysRevE.90.012120} {\bibfield  {journal} {\bibinfo
			{journal} {Phys. Rev. E}\ }\textbf {\bibinfo {volume} {90}},\ \bibinfo
		{pages} {012120} (\bibinfo {year} {2014})}\BibitemShut {NoStop}%
	\bibitem [{\citenamefont {Thewes}\ and\ \citenamefont
		{Fernandes}(2020)}]{PhysRevE.101.062138}%
	\BibitemOpen
	\bibfield  {author} {\bibinfo {author} {\bibfnamefont {F.~C.}\ \bibnamefont
			{Thewes}}\ and\ \bibinfo {author} {\bibfnamefont {H.~C.~M.}\ \bibnamefont
			{Fernandes}},\ }\bibfield  {title} {\bibinfo {title} {Phase transitions in
			hard-core lattice gases on the honeycomb lattice},\ }\href
	{https://doi.org/10.1103/PhysRevE.101.062138} {\bibfield  {journal} {\bibinfo
			{journal} {Phys. Rev. E}\ }\textbf {\bibinfo {volume} {101}},\ \bibinfo
		{pages} {062138} (\bibinfo {year} {2020})}\BibitemShut {NoStop}%
	\bibitem [{\citenamefont {Akimenko}\ \emph {et~al.}(2019)\citenamefont
		{Akimenko}, \citenamefont {Gorbunov}, \citenamefont {Myshlyavtsev},\ and\
		\citenamefont {Stishenko}}]{akimenko2019tensor}%
	\BibitemOpen
	\bibfield  {author} {\bibinfo {author} {\bibfnamefont {S.~S.}\ \bibnamefont
			{Akimenko}}, \bibinfo {author} {\bibfnamefont {V.~A.}\ \bibnamefont
			{Gorbunov}}, \bibinfo {author} {\bibfnamefont {A.~V.}\ \bibnamefont
			{Myshlyavtsev}},\ and\ \bibinfo {author} {\bibfnamefont {P.~V.}\ \bibnamefont
			{Stishenko}},\ }\bibfield  {title} {\bibinfo {title} {Tensor renormalization
			group study of hard-disk models on a triangular lattice},\ }\href@noop {}
	{\bibfield  {journal} {\bibinfo  {journal} {Phys. Rev. E}\ }\textbf {\bibinfo
			{volume} {100}},\ \bibinfo {pages} {022108} (\bibinfo {year}
		{2019})}\BibitemShut {NoStop}%
	\bibitem [{\citenamefont {Domb}(1958)}]{1958-d-nc-theoretical}%
	\BibitemOpen
	\bibfield  {author} {\bibinfo {author} {\bibfnamefont {C.}~\bibnamefont
			{Domb}},\ }\bibfield  {title} {\bibinfo {title} {Some theoretical aspects of
			melting},\ }\href {https://doi.org/10.1007/BF02824224} {\bibfield  {journal}
		{\bibinfo  {journal} {Il Nuovo Cimento (1955-1965)}\ }\textbf {\bibinfo
			{volume} {9}},\ \bibinfo {pages} {9} (\bibinfo {year} {1958})}\BibitemShut
	{NoStop}%
	\bibitem [{\citenamefont {Burley}(1960)}]{1960-b-pps-lattice}%
	\BibitemOpen
	\bibfield  {author} {\bibinfo {author} {\bibfnamefont {D.~M.}\ \bibnamefont
			{Burley}},\ }\bibfield  {title} {\bibinfo {title} {A lattice model of a
			classical hard sphere gas},\ }\href
	{http://stacks.iop.org/0370-1328/75/i=2/a=313} {\bibfield  {journal}
		{\bibinfo  {journal} {Proc. Phys. Soc.}\ }\textbf {\bibinfo {volume} {75}},\
		\bibinfo {pages} {262} (\bibinfo {year} {1960})}\BibitemShut {NoStop}%
	\bibitem [{\citenamefont {Bellemans}\ and\ \citenamefont
		{Nigam}(1967)}]{1967-bn-jcp-phase}%
	\BibitemOpen
	\bibfield  {author} {\bibinfo {author} {\bibfnamefont {A.}~\bibnamefont
			{Bellemans}}\ and\ \bibinfo {author} {\bibfnamefont {R.~K.}\ \bibnamefont
			{Nigam}},\ }\bibfield  {title} {\bibinfo {title} {Phase transitions in
			two‐dimensional lattice gases of hard‐square molecules},\ }\href
	{https://doi.org/10.1063/1.1841157} {\bibfield  {journal} {\bibinfo
			{journal} {J. Chem. Phys.}\ }\textbf {\bibinfo {volume} {46}},\ \bibinfo
		{pages} {2922} (\bibinfo {year} {1967})}\BibitemShut {NoStop}%
	\bibitem [{\citenamefont {Bellemans}\ and\ \citenamefont
		{Nigam}(1966)}]{1966-bn-prl-phase}%
	\BibitemOpen
	\bibfield  {author} {\bibinfo {author} {\bibfnamefont {A.}~\bibnamefont
			{Bellemans}}\ and\ \bibinfo {author} {\bibfnamefont {R.~K.}\ \bibnamefont
			{Nigam}},\ }\bibfield  {title} {\bibinfo {title} {Phase transitions in the
			hard-square lattice gas},\ }\href
	{https://doi.org/10.1103/PhysRevLett.16.1038} {\bibfield  {journal} {\bibinfo
			{journal} {Phys. Rev. Lett.}\ }\textbf {\bibinfo {volume} {16}},\ \bibinfo
		{pages} {1038} (\bibinfo {year} {1966})}\BibitemShut {NoStop}%
	\bibitem [{\citenamefont {Kasteleyn}(1961)}]{1961-k-physica-statistics}%
	\BibitemOpen
	\bibfield  {author} {\bibinfo {author} {\bibfnamefont {P.}~\bibnamefont
			{Kasteleyn}},\ }\bibfield  {title} {\bibinfo {title} {The statistics of
			dimers on a lattice},\ }\href
	{https://doi.org/http://dx.doi.org/10.1016/0031-8914(61)90063-5} {\bibfield
		{journal} {\bibinfo  {journal} {Physica}\ }\textbf {\bibinfo {volume} {27}},\
		\bibinfo {pages} {1209 } (\bibinfo {year} {1961})}\BibitemShut {NoStop}%
	\bibitem [{\citenamefont {Kundu}\ \emph {et~al.}(2012)\citenamefont {Kundu},
		\citenamefont {Rajesh}, \citenamefont {Dhar},\ and\ \citenamefont
		{Stilck}}]{2012-krds-aipcp-monte}%
	\BibitemOpen
	\bibfield  {author} {\bibinfo {author} {\bibfnamefont {J.}~\bibnamefont
			{Kundu}}, \bibinfo {author} {\bibfnamefont {R.}~\bibnamefont {Rajesh}},
		\bibinfo {author} {\bibfnamefont {D.}~\bibnamefont {Dhar}},\ and\ \bibinfo
		{author} {\bibfnamefont {J.~F.}\ \bibnamefont {Stilck}},\ }\bibfield  {title}
	{\bibinfo {title} {A monte carlo algorithm for studying phase transition in
			systems of hard rigid rods},\ }\href {https://doi.org/10.1063/1.4709907}
	{\bibfield  {journal} {\bibinfo  {journal} {AIP Conf. Proc.}\ }\textbf
		{\bibinfo {volume} {1447}},\ \bibinfo {pages} {113} (\bibinfo {year}
		{2012})}\BibitemShut {NoStop}%
	\bibitem [{\citenamefont {Ramola}\ \emph {et~al.}(2015)\citenamefont {Ramola},
		\citenamefont {Damle},\ and\ \citenamefont {Dhar}}]{2015-rdd-prl-columnar}%
	\BibitemOpen
	\bibfield  {author} {\bibinfo {author} {\bibfnamefont {K.}~\bibnamefont
			{Ramola}}, \bibinfo {author} {\bibfnamefont {K.}~\bibnamefont {Damle}},\ and\
		\bibinfo {author} {\bibfnamefont {D.}~\bibnamefont {Dhar}},\ }\bibfield
	{title} {\bibinfo {title} {Columnar order and ashkin-teller criticality in
			mixtures of hard squares and dimers},\ }\href
	{https://doi.org/10.1103/PhysRevLett.114.190601} {\bibfield  {journal}
		{\bibinfo  {journal} {Phys. Rev. Lett.}\ }\textbf {\bibinfo {volume} {114}},\
		\bibinfo {pages} {190601} (\bibinfo {year} {2015})}\BibitemShut {NoStop}%
	\bibitem [{\citenamefont {Berg}\ and\ \citenamefont
		{Neuhaus}(1992)}]{berg1992multicanonical}%
	\BibitemOpen
	\bibfield  {author} {\bibinfo {author} {\bibfnamefont {B.~A.}\ \bibnamefont
			{Berg}}\ and\ \bibinfo {author} {\bibfnamefont {T.}~\bibnamefont {Neuhaus}},\
	}\bibfield  {title} {\bibinfo {title} {Multicanonical ensemble: A new
			approach to simulate first-order phase transitions},\ }\href@noop {}
	{\bibfield  {journal} {\bibinfo  {journal} {Phys. Rev. Lett.}\ }\textbf
		{\bibinfo {volume} {68}},\ \bibinfo {pages} {9} (\bibinfo {year}
		{1992})}\BibitemShut {NoStop}%
	\bibitem [{\citenamefont {Lee}(1993)}]{lee1993new}%
	\BibitemOpen
	\bibfield  {author} {\bibinfo {author} {\bibfnamefont {J.}~\bibnamefont
			{Lee}},\ }\bibfield  {title} {\bibinfo {title} {New monte carlo algorithm:
			entropic sampling},\ }\href@noop {} {\bibfield  {journal} {\bibinfo
			{journal} {Phys. Rev. Lett.}\ }\textbf {\bibinfo {volume} {71}},\ \bibinfo
		{pages} {211} (\bibinfo {year} {1993})}\BibitemShut {NoStop}%
	\bibitem [{\citenamefont {de~oliveira}\ \emph {et~al.}(1996)\citenamefont
		{de~oliveira}, \citenamefont {Penna},\ and\ \citenamefont
		{Herrmann}}]{herrmann1996broad}%
	\BibitemOpen
	\bibfield  {author} {\bibinfo {author} {\bibfnamefont {P.}~\bibnamefont
			{de~oliveira}}, \bibinfo {author} {\bibfnamefont {T.}~\bibnamefont {Penna}},\
		and\ \bibinfo {author} {\bibfnamefont {H.}~\bibnamefont {Herrmann}},\
	}\bibfield  {title} {\bibinfo {title} {Broad histogram method},\ }\href@noop
	{} {\bibfield  {journal} {\bibinfo  {journal} {Braz. J. Phys.}\ }\textbf
		{\bibinfo {volume} {26}} ,\ \bibinfo
		{pages} {677} (\bibinfo {year} {1996})}\BibitemShut {NoStop}%
	\bibitem [{\citenamefont {Wang}\ and\ \citenamefont
		{Lee}(2000)}]{wang2000monte}%
	\BibitemOpen
	\bibfield  {author} {\bibinfo {author} {\bibfnamefont {J.-S.}\ \bibnamefont
			{Wang}}\ and\ \bibinfo {author} {\bibfnamefont {L.~W.}\ \bibnamefont {Lee}},\
	}\bibfield  {title} {\bibinfo {title} {Monte carlo algorithms based on the
			number of potential moves},\ }\href@noop {} {\bibfield  {journal} {\bibinfo
			{journal} {Comput. Phys. Commun.}\ }\textbf {\bibinfo {volume} {127}},\
		\bibinfo {pages} {131} (\bibinfo {year} {2000})}\BibitemShut {NoStop}%
	\bibitem [{\citenamefont {Wang}\ and\ \citenamefont
		{Landau}(2001{\natexlab{a}})}]{2001-wl-prl-efficient}%
	\BibitemOpen
	\bibfield  {author} {\bibinfo {author} {\bibfnamefont {F.}~\bibnamefont
			{Wang}}\ and\ \bibinfo {author} {\bibfnamefont {D.~P.}\ \bibnamefont
			{Landau}},\ }\bibfield  {title} {\bibinfo {title} {Efficient, multiple-range
			random walk algorithm to calculate the density of states},\ }\href
	{https://doi.org/10.1103/PhysRevLett.86.2050} {\bibfield  {journal} {\bibinfo
			{journal} {Phys. Rev. Lett.}\ }\textbf {\bibinfo {volume} {86}},\ \bibinfo
		{pages} {2050} (\bibinfo {year} {2001}{\natexlab{a}})}\BibitemShut {NoStop}%
	\bibitem [{\citenamefont {Wang}\ and\ \citenamefont
		{Landau}(2001{\natexlab{b}})}]{2001-wl-pre-determining}%
	\BibitemOpen
	\bibfield  {author} {\bibinfo {author} {\bibfnamefont {F.}~\bibnamefont
			{Wang}}\ and\ \bibinfo {author} {\bibfnamefont {D.~P.}\ \bibnamefont
			{Landau}},\ }\bibfield  {title} {\bibinfo {title} {Determining the density of
			states for classical statistical models: A random walk algorithm to produce a
			flat histogram},\ }\href {https://doi.org/10.1103/PhysRevE.64.056101}
	{\bibfield  {journal} {\bibinfo  {journal} {Phys. Rev. E}\ }\textbf {\bibinfo
			{volume} {64}},\ \bibinfo {pages} {056101} (\bibinfo {year}
		{2001}{\natexlab{b}})}\BibitemShut {NoStop}%
	\bibitem [{\citenamefont {Zhou}\ and\ \citenamefont
		{Bhatt}(2005)}]{zhou2005understanding}%
	\BibitemOpen
	\bibfield  {author} {\bibinfo {author} {\bibfnamefont {C.}~\bibnamefont
			{Zhou}}\ and\ \bibinfo {author} {\bibfnamefont {R.}~\bibnamefont {Bhatt}},\
	}\bibfield  {title} {\bibinfo {title} {Understanding and improving the
			wang-landau algorithm},\ }\href@noop {} {\bibfield  {journal} {\bibinfo
			{journal} {Phys. Rev. E}\ }\textbf {\bibinfo {volume} {72}},\ \bibinfo
		{pages} {025701} (\bibinfo {year} {2005})}\BibitemShut {NoStop}%
	\bibitem [{\citenamefont {Singh}\ \emph {et~al.}(2012)\citenamefont {Singh},
		\citenamefont {Chopra},\ and\ \citenamefont {de~Pablo}}]{singh2012density}%
	\BibitemOpen
	\bibfield  {author} {\bibinfo {author} {\bibfnamefont {S.}~\bibnamefont
			{Singh}}, \bibinfo {author} {\bibfnamefont {M.}~\bibnamefont {Chopra}},\ and\
		\bibinfo {author} {\bibfnamefont {J.~J.}\ \bibnamefont {de~Pablo}},\
	}\bibfield  {title} {\bibinfo {title} {Density of states--based molecular
			simulations},\ }\href@noop {} {\bibfield  {journal} {\bibinfo  {journal}
			{Annu. Rev. Chem. Biomol. Eng.}\ }\textbf {\bibinfo {volume} {3}},\ \bibinfo
		{pages} {369} (\bibinfo {year} {2012})}\BibitemShut {NoStop}%
	\bibitem [{\citenamefont {Cunha-Netto}\ and\ \citenamefont
		{Dickman}(2011)}]{cunha2011critical}%
	\BibitemOpen
	\bibfield  {author} {\bibinfo {author} {\bibfnamefont {A.}~\bibnamefont
			{Cunha-Netto}}\ and\ \bibinfo {author} {\bibfnamefont {R.}~\bibnamefont
			{Dickman}},\ }\bibfield  {title} {\bibinfo {title} {Critical behavior of
			hard-core lattice gases: Wang--landau sampling with adaptive windows},\
	}\href@noop {} {\bibfield  {journal} {\bibinfo  {journal} {Comput. Phys.
				Commun.}\ }\textbf {\bibinfo {volume} {182}},\ \bibinfo {pages} {719}
		(\bibinfo {year} {2011})}\BibitemShut {NoStop}%
	\bibitem [{\citenamefont {Belardinelli}\ and\ \citenamefont
		{Pereyra}(2007{\natexlab{a}})}]{belardinelli2007fast}%
	\BibitemOpen
	\bibfield  {author} {\bibinfo {author} {\bibfnamefont {R.}~\bibnamefont
			{Belardinelli}}\ and\ \bibinfo {author} {\bibfnamefont {V.}~\bibnamefont
			{Pereyra}},\ }\bibfield  {title} {\bibinfo {title} {Fast algorithm to
			calculate density of states},\ }\href@noop {} {\bibfield  {journal} {\bibinfo
			{journal} {Phys. Rev. E}\ }\textbf {\bibinfo {volume} {75}},\ \bibinfo
		{pages} {046701} (\bibinfo {year} {2007}{\natexlab{a}})}\BibitemShut
	{NoStop}%
	\bibitem [{\citenamefont {Belardinelli}\ and\ \citenamefont
		{Pereyra}(2007{\natexlab{b}})}]{belardinelli2007wang}%
	\BibitemOpen
	\bibfield  {author} {\bibinfo {author} {\bibfnamefont {R.}~\bibnamefont
			{Belardinelli}}\ and\ \bibinfo {author} {\bibfnamefont {V.}~\bibnamefont
			{Pereyra}},\ }\bibfield  {title} {\bibinfo {title} {Wang-landau algorithm: A
			theoretical analysis of the saturation of the error},\ }\href@noop {}
	{\bibfield  {journal} {\bibinfo  {journal} {J. Chem. Phys.}\ }\textbf
		{\bibinfo {volume} {127}},\ \bibinfo {pages} {184105} (\bibinfo {year}
		{2007}{\natexlab{b}})}\BibitemShut {NoStop}%
	\bibitem [{\citenamefont {Dickman}\ and\ \citenamefont
		{Cunha-Netto}(2011)}]{2011-dc-pre-complete}%
	\BibitemOpen
	\bibfield  {author} {\bibinfo {author} {\bibfnamefont {R.}~\bibnamefont
			{Dickman}}\ and\ \bibinfo {author} {\bibfnamefont {A.~G.}\ \bibnamefont
			{Cunha-Netto}},\ }\bibfield  {title} {\bibinfo {title} {Complete
			high-precision entropic sampling},\ }\href
	{https://doi.org/10.1103/PhysRevE.84.026701} {\bibfield  {journal} {\bibinfo
			{journal} {Phys. Rev. E}\ }\textbf {\bibinfo {volume} {84}},\ \bibinfo
		{pages} {026701} (\bibinfo {year} {2011})}\BibitemShut {NoStop}%
	\bibitem [{\citenamefont {Belardinelli}\ \emph {et~al.}(2014)\citenamefont
		{Belardinelli}, \citenamefont {Pereyra}, \citenamefont {Dickman},\ and\
		\citenamefont {Lourenço}}]{2014-bpdl-jsm-intrinsic}%
	\BibitemOpen
	\bibfield  {author} {\bibinfo {author} {\bibfnamefont {R.~E.}\ \bibnamefont
			{Belardinelli}}, \bibinfo {author} {\bibfnamefont {V.~D.}\ \bibnamefont
			{Pereyra}}, \bibinfo {author} {\bibfnamefont {R.}~\bibnamefont {Dickman}},\
		and\ \bibinfo {author} {\bibfnamefont {B.~J.}\ \bibnamefont {Lourenço}},\
	}\bibfield  {title} {\bibinfo {title} {Intrinsic convergence properties of
			entropic sampling algorithms},\ }\href
	{http://stacks.iop.org/1742-5468/2014/i=7/a=P07007} {\bibfield  {journal}
		{\bibinfo  {journal} {J. Stat. Mech.}\ }\textbf {\bibinfo {volume} {2014}},\
		\bibinfo {pages} {P07007} (\bibinfo {year} {2014})}\BibitemShut {NoStop}%
	\bibitem [{\citenamefont {Darjani}\ \emph {et~al.}(2017)\citenamefont
		{Darjani}, \citenamefont {Koplik},\ and\ \citenamefont
		{Pauchard}}]{darjani2017extracting}%
	\BibitemOpen
	\bibfield  {author} {\bibinfo {author} {\bibfnamefont {S.}~\bibnamefont
			{Darjani}}, \bibinfo {author} {\bibfnamefont {J.}~\bibnamefont {Koplik}},\
		and\ \bibinfo {author} {\bibfnamefont {V.}~\bibnamefont {Pauchard}},\
	}\bibfield  {title} {\bibinfo {title} {Extracting the equation of state of
			lattice gases from random sequential adsorption simulations by means of the
			gibbs adsorption isotherm},\ }\href@noop {} {\bibfield  {journal} {\bibinfo
			{journal} {Phys. Rev. E}\ }\textbf {\bibinfo {volume} {96}},\ \bibinfo
		{pages} {052803} (\bibinfo {year} {2017})}\BibitemShut {NoStop}%
	\bibitem [{\citenamefont {Darjani}\ \emph {et~al.}(2019)\citenamefont
		{Darjani}, \citenamefont {Koplik}, \citenamefont {Banerjee},\ and\
		\citenamefont {Pauchard}}]{darjani2019liquid}%
	\BibitemOpen
	\bibfield  {author} {\bibinfo {author} {\bibfnamefont {S.}~\bibnamefont
			{Darjani}}, \bibinfo {author} {\bibfnamefont {J.}~\bibnamefont {Koplik}},
		\bibinfo {author} {\bibfnamefont {S.}~\bibnamefont {Banerjee}},\ and\
		\bibinfo {author} {\bibfnamefont {V.}~\bibnamefont {Pauchard}},\ }\bibfield
	{title} {\bibinfo {title} {Liquid-hexatic-solid phase transition of a
			hard-core lattice gas with third neighbor exclusion},\ }\href@noop {}
	{\bibfield  {journal} {\bibinfo  {journal} {J. Chem. Phys.}\ }\textbf
		{\bibinfo {volume} {151}},\ \bibinfo {pages} {104702} (\bibinfo {year}
		{2019})}\BibitemShut {NoStop}%
	\bibitem [{\citenamefont {Fisher}(1967)}]{fisher1967theory}%
	\BibitemOpen
	\bibfield  {author} {\bibinfo {author} {\bibfnamefont {M.~E.}\ \bibnamefont
			{Fisher}},\ }\bibfield  {title} {\bibinfo {title} {The theory of equilibrium
			critical phenomena},\ }\href@noop {} {\bibfield  {journal} {\bibinfo
			{journal} {Rep. Prog. Phys.}\ }\textbf {\bibinfo {volume} {30}},\ \bibinfo
		{pages} {615} (\bibinfo {year} {1967})}\BibitemShut {NoStop}%
	\bibitem [{\citenamefont {Fisher}\ and\ \citenamefont
		{Barber}(1972)}]{fisher1972scaling}%
	\BibitemOpen
	\bibfield  {author} {\bibinfo {author} {\bibfnamefont {M.~E.}\ \bibnamefont
			{Fisher}}\ and\ \bibinfo {author} {\bibfnamefont {M.~N.}\ \bibnamefont
			{Barber}},\ }\bibfield  {title} {\bibinfo {title} {Scaling theory for
			finite-size effects in the critical region},\ }\href@noop {} {\bibfield
		{journal} {\bibinfo  {journal} {Phys. Rev. Lett.}\ }\textbf {\bibinfo
			{volume} {28}},\ \bibinfo {pages} {1516} (\bibinfo {year}
		{1972})}\BibitemShut {NoStop}%
	\bibitem [{\citenamefont {Fisher}(1983)}]{fisher1983scaling}%
	\BibitemOpen
	\bibfield  {author} {\bibinfo {author} {\bibfnamefont {M.~E.}\ \bibnamefont
			{Fisher}},\ }\bibfield  {title} {\bibinfo {title} {Scaling, universality and
			renormalization group theory},\ }in\ \href@noop {} {\emph {\bibinfo
			{booktitle} {Critical phenomena}}}\ (\bibinfo  {publisher} {Springer},\
	\bibinfo {year} {1983})\ pp.\ \bibinfo {pages} {1--139}\BibitemShut {NoStop}%
	\bibitem [{\citenamefont {Pelissetto}\ and\ \citenamefont
		{Vicari}(2002)}]{pelissetto2002critical}%
	\BibitemOpen
	\bibfield  {author} {\bibinfo {author} {\bibfnamefont {A.}~\bibnamefont
			{Pelissetto}}\ and\ \bibinfo {author} {\bibfnamefont {E.}~\bibnamefont
			{Vicari}},\ }\bibfield  {title} {\bibinfo {title} {Critical phenomena and
			renormalization-group theory},\ }\href@noop {} {\bibfield  {journal}
		{\bibinfo  {journal} {Phys. Rep.}\ }\textbf {\bibinfo {volume} {368}},\
		\bibinfo {pages} {549} (\bibinfo {year} {2002})}\BibitemShut {NoStop}%
	\bibitem [{\citenamefont {Rodrigues}\ and\ \citenamefont
		{Oliveira}(2021)}]{rodrigues2021husimi}%
	\BibitemOpen
	\bibfield  {author} {\bibinfo {author} {\bibfnamefont {N.~T.}\ \bibnamefont
			{Rodrigues}}\ and\ \bibinfo {author} {\bibfnamefont {T.~J.}\ \bibnamefont
			{Oliveira}},\ }\bibfield  {title} {\bibinfo {title} {Husimi-lattice solutions
			and the coherent-anomaly-method analysis for hard-square lattice gases},\
	}\href@noop {} {\bibfield  {journal} {\bibinfo  {journal} {Phys. Rev. E}\
		}\textbf {\bibinfo {volume} {103}},\ \bibinfo {pages} {032153} (\bibinfo
		{year} {2021})}\BibitemShut {NoStop}%
	\bibitem [{\citenamefont {Guo}\ and\ \citenamefont
		{Bl\"ote}(2002)}]{2002-gb-pre-finite}%
	\BibitemOpen
	\bibfield  {author} {\bibinfo {author} {\bibfnamefont {W.}~\bibnamefont
			{Guo}}\ and\ \bibinfo {author} {\bibfnamefont {H.~W.~J.}\ \bibnamefont
			{Bl\"ote}},\ }\bibfield  {title} {\bibinfo {title} {Finite-size analysis of
			the hard-square lattice gas},\ }\href
	{https://doi.org/10.1103/PhysRevE.66.046140} {\bibfield  {journal} {\bibinfo
			{journal} {Phys. Rev. E}\ }\textbf {\bibinfo {volume} {66}},\ \bibinfo
		{pages} {046140} (\bibinfo {year} {2002})}\BibitemShut {NoStop}%
	\bibitem [{\citenamefont {Zhitomirsky}\ and\ \citenamefont
		{Tsunetsugu}(2007)}]{2007-zt-prb-lattice}%
	\BibitemOpen
	\bibfield  {author} {\bibinfo {author} {\bibfnamefont {M.~E.}\ \bibnamefont
			{Zhitomirsky}}\ and\ \bibinfo {author} {\bibfnamefont {H.}~\bibnamefont
			{Tsunetsugu}},\ }\bibfield  {title} {\bibinfo {title} {Lattice gas
			description of pyrochlore and checkerboard antiferromagnets in a strong
			magnetic field},\ }\href {https://doi.org/10.1103/PhysRevB.75.224416}
	{\bibfield  {journal} {\bibinfo  {journal} {Phys. Rev. B}\ }\textbf {\bibinfo
			{volume} {75}},\ \bibinfo {pages} {224416} (\bibinfo {year}
		{2007})}\BibitemShut {NoStop}%
	\bibitem [{\citenamefont {Feng}\ \emph {et~al.}(2011)\citenamefont {Feng},
		\citenamefont {Bl\"ote},\ and\ \citenamefont
		{Nienhuis}}]{2011-fbn-pre-lattice}%
	\BibitemOpen
	\bibfield  {author} {\bibinfo {author} {\bibfnamefont {X.}~\bibnamefont
			{Feng}}, \bibinfo {author} {\bibfnamefont {H.~W.~J.}\ \bibnamefont
			{Bl\"ote}},\ and\ \bibinfo {author} {\bibfnamefont {B.}~\bibnamefont
			{Nienhuis}},\ }\bibfield  {title} {\bibinfo {title} {Lattice gas with
			nearest- and next-to-nearest-neighbor exclusion},\ }\href
	{https://doi.org/10.1103/PhysRevE.83.061153} {\bibfield  {journal} {\bibinfo
			{journal} {Phys. Rev. E}\ }\textbf {\bibinfo {volume} {83}},\ \bibinfo
		{pages} {061153} (\bibinfo {year} {2011})}\BibitemShut {NoStop}%
	\bibitem [{\citenamefont {Amar}\ \emph {et~al.}(1984)\citenamefont {Amar},
		\citenamefont {Kaski},\ and\ \citenamefont {Gunton}}]{1984-akg-prb-square}%
	\BibitemOpen
	\bibfield  {author} {\bibinfo {author} {\bibfnamefont {J.}~\bibnamefont
			{Amar}}, \bibinfo {author} {\bibfnamefont {K.}~\bibnamefont {Kaski}},\ and\
		\bibinfo {author} {\bibfnamefont {J.~D.}\ \bibnamefont {Gunton}},\ }\bibfield
	{title} {\bibinfo {title} {Square-lattice-gas model with repulsive nearest-
			and next-nearest-neighbor interactions},\ }\href
	{https://doi.org/10.1103/PhysRevB.29.1462} {\bibfield  {journal} {\bibinfo
			{journal} {Phys. Rev. B}\ }\textbf {\bibinfo {volume} {29}},\ \bibinfo
		{pages} {1462} (\bibinfo {year} {1984})}\BibitemShut {NoStop}%
	\bibitem [{\citenamefont {Ramola}\ and\ \citenamefont
		{Dhar}(2012)}]{2012-rd-pre-high}%
	\BibitemOpen
	\bibfield  {author} {\bibinfo {author} {\bibfnamefont {K.}~\bibnamefont
			{Ramola}}\ and\ \bibinfo {author} {\bibfnamefont {D.}~\bibnamefont {Dhar}},\
	}\bibfield  {title} {\bibinfo {title} {High-activity perturbation expansion
			for the hard square lattice gas},\ }\href
	{https://doi.org/10.1103/PhysRevE.86.031135} {\bibfield  {journal} {\bibinfo
			{journal} {Phys. Rev. E}\ }\textbf {\bibinfo {volume} {86}},\ \bibinfo
		{pages} {031135} (\bibinfo {year} {2012})}\BibitemShut {NoStop}%
	\bibitem [{\citenamefont {Nath}\ \emph {et~al.}(2015)\citenamefont {Nath},
		\citenamefont {Kundu},\ and\ \citenamefont {Rajesh}}]{2015-nkr-jsp-high}%
	\BibitemOpen
	\bibfield  {author} {\bibinfo {author} {\bibfnamefont {T.}~\bibnamefont
			{Nath}}, \bibinfo {author} {\bibfnamefont {J.}~\bibnamefont {Kundu}},\ and\
		\bibinfo {author} {\bibfnamefont {R.}~\bibnamefont {Rajesh}},\ }\bibfield
	{title} {\bibinfo {title} {High-activity expansion for the columnar phase of
			the hard rectangle gas},\ }\href {https://doi.org/10.1007/s10955-015-1285-y}
	{\bibfield  {journal} {\bibinfo  {journal} {J. Stat. Phys.}\ }\textbf
		{\bibinfo {volume} {160}},\ \bibinfo {pages} {1173} (\bibinfo {year}
		{2015})}\BibitemShut {NoStop}%
	\bibitem [{\citenamefont {Slotte}(1983)}]{1983-s-jpc-phase}%
	\BibitemOpen
	\bibfield  {author} {\bibinfo {author} {\bibfnamefont {P.~A.}\ \bibnamefont
			{Slotte}},\ }\bibfield  {title} {\bibinfo {title} {Phase diagram of the
			square-lattice ising model with first- and second-neighbour interactions},\
	}\href {http://stacks.iop.org/0022-3719/16/i=15/a=015} {\bibfield  {journal}
		{\bibinfo  {journal} {J. Phys., C, Solid State Phys.}\ }\textbf {\bibinfo
			{volume} {16}},\ \bibinfo {pages} {2935} (\bibinfo {year}
		{1983})}\BibitemShut {NoStop}%
	\bibitem [{\citenamefont {Nath}\ \emph {et~al.}(2016)\citenamefont {Nath},
		\citenamefont {Dhar},\ and\ \citenamefont {Rajesh}}]{2016-ndr-epl-stability}%
	\BibitemOpen
	\bibfield  {author} {\bibinfo {author} {\bibfnamefont {T.}~\bibnamefont
			{Nath}}, \bibinfo {author} {\bibfnamefont {D.}~\bibnamefont {Dhar}},\ and\
		\bibinfo {author} {\bibfnamefont {R.}~\bibnamefont {Rajesh}},\ }\bibfield
	{title} {\bibinfo {title} {Stability of columnar order in assemblies of hard
			rectangles or squares},\ }\href
	{http://stacks.iop.org/0295-5075/114/i=1/a=10003} {\bibfield  {journal}
		{\bibinfo  {journal} {Eur. Phys. Lett.}\ }\textbf {\bibinfo {volume} {114}},\
		\bibinfo {pages} {10003} (\bibinfo {year} {2016})}\BibitemShut {NoStop}%
	\bibitem [{\citenamefont {Mandal}\ \emph {et~al.}(2017)\citenamefont {Mandal},
		\citenamefont {Nath},\ and\ \citenamefont {Rajesh}}]{mandal2017estimating}%
	\BibitemOpen
	\bibfield  {author} {\bibinfo {author} {\bibfnamefont {D.}~\bibnamefont
			{Mandal}}, \bibinfo {author} {\bibfnamefont {T.}~\bibnamefont {Nath}},\ and\
		\bibinfo {author} {\bibfnamefont {R.}~\bibnamefont {Rajesh}},\ }\bibfield
	{title} {\bibinfo {title} {Estimating the critical parameters of the hard
			square lattice gas model},\ }\href@noop {} {\bibfield  {journal} {\bibinfo
			{journal} {J. Stat. Mech.}\ }\textbf {\bibinfo {volume} {2017}},\ \bibinfo
		{pages} {043201} (\bibinfo {year} {2017})}\BibitemShut {NoStop}%
	\bibitem [{\citenamefont {Fiore}\ and\ \citenamefont
		{da~Luz}(2013)}]{2013-fl-jcp-exploiting}%
	\BibitemOpen
	\bibfield  {author} {\bibinfo {author} {\bibfnamefont {C.~E.}\ \bibnamefont
			{Fiore}}\ and\ \bibinfo {author} {\bibfnamefont {M.~G.~E.}\ \bibnamefont
			{da~Luz}},\ }\bibfield  {title} {\bibinfo {title} {Exploiting a semi-analytic
			approach to study first order phase transitions},\ }\href@noop {} {\bibfield
		{journal} {\bibinfo  {journal} {J. Chem. Phys.}\ }\textbf {\bibinfo {volume}
			{138}},\ \bibinfo {pages} {014105} (\bibinfo {year} {2013})}\BibitemShut
	{NoStop}%
	\bibitem [{\citenamefont {Eisenberg}\ and\ \citenamefont
		{Baram}(2005)}]{2005-eb-epl-first}%
	\BibitemOpen
	\bibfield  {author} {\bibinfo {author} {\bibfnamefont {E.}~\bibnamefont
			{Eisenberg}}\ and\ \bibinfo {author} {\bibfnamefont {A.}~\bibnamefont
			{Baram}},\ }\bibfield  {title} {\bibinfo {title} {A first-order phase
			transition and a super-cooled fluid in a two-dimensional lattice gas model},\
	}\href {http://stacks.iop.org/0295-5075/71/i=6/a=900} {\bibfield  {journal}
		{\bibinfo  {journal} {Eur. Phys. Lett.}\ }\textbf {\bibinfo {volume} {71}},\
		\bibinfo {pages} {900} (\bibinfo {year} {2005})}\BibitemShut {NoStop}%
	\bibitem [{\citenamefont {Orban}\ and\ \citenamefont
		{Belle}(1982)}]{1982-ob-jpa-phase}%
	\BibitemOpen
	\bibfield  {author} {\bibinfo {author} {\bibfnamefont {J.}~\bibnamefont
			{Orban}}\ and\ \bibinfo {author} {\bibfnamefont {D.~V.}\ \bibnamefont
			{Belle}},\ }\bibfield  {title} {\bibinfo {title} {Phase transition in a
			lattice gas with extended hard core},\ }\href
	{http://stacks.iop.org/0305-4470/15/i=9/a=012} {\bibfield  {journal}
		{\bibinfo  {journal} {J. Phys. A Math. Theor.}\ }\textbf {\bibinfo {volume}
			{15}},\ \bibinfo {pages} {L501} (\bibinfo {year} {1982})}\BibitemShut
	{NoStop}%
	\bibitem [{\citenamefont {Eisenberg}\ and\ \citenamefont
		{Baram}(2000)}]{2000-eb-jpa-random}%
	\BibitemOpen
	\bibfield  {author} {\bibinfo {author} {\bibfnamefont {E.}~\bibnamefont
			{Eisenberg}}\ and\ \bibinfo {author} {\bibfnamefont {A.}~\bibnamefont
			{Baram}},\ }\bibfield  {title} {\bibinfo {title} {Random closest packing in a
			2d lattice model},\ }\href {http://stacks.iop.org/0305-4470/33/i=9/a=302}
	{\bibfield  {journal} {\bibinfo  {journal} {J. Phys. A}\ }\textbf {\bibinfo
			{volume} {33}},\ \bibinfo {pages} {1729} (\bibinfo {year}
		{2000})}\BibitemShut {NoStop}%
	\bibitem [{\citenamefont {Rotman}\ and\ \citenamefont
		{Eisenberg}(2009)}]{rotman2009ideal}%
	\BibitemOpen
	\bibfield  {author} {\bibinfo {author} {\bibfnamefont {Z.}~\bibnamefont
			{Rotman}}\ and\ \bibinfo {author} {\bibfnamefont {E.}~\bibnamefont
			{Eisenberg}},\ }\bibfield  {title} {\bibinfo {title} {Ideal glass transition
			in a simple two-dimensional lattice model},\ }\href@noop {} {\bibfield
		{journal} {\bibinfo  {journal} {Phys. Rev. E}\ }\textbf {\bibinfo {volume}
			{80}},\ \bibinfo {pages} {060104} (\bibinfo {year} {2009})}\BibitemShut
	{NoStop}%
	\bibitem [{\citenamefont {Rotman}\ and\ \citenamefont
		{Eisenberg}(2010)}]{rotman2010direct}%
	\BibitemOpen
	\bibfield  {author} {\bibinfo {author} {\bibfnamefont {Z.}~\bibnamefont
			{Rotman}}\ and\ \bibinfo {author} {\bibfnamefont {E.}~\bibnamefont
			{Eisenberg}},\ }\bibfield  {title} {\bibinfo {title} {Direct measurements of
			the dynamical correlation length indicate its divergence at an athermal glass
			transition},\ }\href@noop {} {\bibfield  {journal} {\bibinfo  {journal}
			{Phys. Rev. Lett.}\ }\textbf {\bibinfo {volume} {105}},\ \bibinfo {pages}
		{225503} (\bibinfo {year} {2010})}\BibitemShut {NoStop}%
	\bibitem [{\citenamefont {Bernard}\ and\ \citenamefont
		{Krauth}(2011)}]{2011-bk-prl-twostep}%
	\BibitemOpen
	\bibfield  {author} {\bibinfo {author} {\bibfnamefont {E.~P.}\ \bibnamefont
			{Bernard}}\ and\ \bibinfo {author} {\bibfnamefont {W.}~\bibnamefont
			{Krauth}},\ }\bibfield  {title} {\bibinfo {title} {Two-step melting in two
			dimensions: First-order liquid-hexatic transition},\ }\href
	{https://doi.org/10.1103/PhysRevLett.107.155704} {\bibfield  {journal}
		{\bibinfo  {journal} {Phys. Rev. Lett.}\ }\textbf {\bibinfo {volume} {107}},\
		\bibinfo {pages} {155704} (\bibinfo {year} {2011})}\BibitemShut {NoStop}%
	\bibitem [{\citenamefont {Heringa}\ and\ \citenamefont
		{Bl{\"o}te}(1996)}]{1996-hb-physica-simple}%
	\BibitemOpen
	\bibfield  {author} {\bibinfo {author} {\bibfnamefont {J.}~\bibnamefont
			{Heringa}}\ and\ \bibinfo {author} {\bibfnamefont {H.}~\bibnamefont
			{Bl{\"o}te}},\ }\bibfield  {title} {\bibinfo {title} {The simple-cubic
			lattice gas with nearest-neighbour exclusion: Ising universality},\ }\href
	{https://doi.org/https://doi.org/10.1016/0378-4371(96)00148-3} {\bibfield
		{journal} {\bibinfo  {journal} {Physica A}\ }\textbf {\bibinfo {volume}
			{232}},\ \bibinfo {pages} {369 } (\bibinfo {year} {1996})}\BibitemShut
	{NoStop}%
	\bibitem [{\citenamefont {Panagiotopoulos}(2005)}]{2005-p-jcp-thermodynamic}%
	\BibitemOpen
	\bibfield  {author} {\bibinfo {author} {\bibfnamefont {A.~Z.}\ \bibnamefont
			{Panagiotopoulos}},\ }\bibfield  {title} {\bibinfo {title} {Thermodynamic
			properties of lattice hard-sphere models},\ }\href
	{https://doi.org/10.1063/1.2008253} {\bibfield  {journal} {\bibinfo
			{journal} {J. Chem. Phys.}\ }\textbf {\bibinfo {volume} {123}},\ \bibinfo
		{pages} {104504} (\bibinfo {year} {2005})}\BibitemShut {NoStop}%
	\bibitem [{\citenamefont {Jaleel}\ \emph {et~al.}(2021)\citenamefont {Jaleel},
		\citenamefont {Mandal},\ and\ \citenamefont {Rajesh}}]{jaleel2021hard}%
	\BibitemOpen
	\bibfield  {author} {\bibinfo {author} {\bibfnamefont {A.~A.~A.}\
			\bibnamefont {Jaleel}}, \bibinfo {author} {\bibfnamefont {D.}~\bibnamefont
			{Mandal}},\ and\ \bibinfo {author} {\bibfnamefont {R.}~\bibnamefont
			{Rajesh}},\ }\bibfield  {title} {\bibinfo {title} {Hard core lattice gas with
			third next-nearest neighbor exclusion on triangular lattice: one or two phase
			transitions?},\ }\href@noop {} {\bibfield  {journal} {\bibinfo  {journal}
			{arXiv:2108.03547}\ } (\bibinfo {year} {2021})}\BibitemShut
	{NoStop}%
	\bibitem [{\citenamefont {Poland}(1983)}]{1983-p-jcp-coexistence}%
	\BibitemOpen
	\bibfield  {author} {\bibinfo {author} {\bibfnamefont {D.}~\bibnamefont
			{Poland}},\ }\bibfield  {title} {\bibinfo {title} {The coexistence curve for
			a mixture of hard‐particle lattice gases},\ }\href
	{https://doi.org/10.1063/1.447023} {\bibfield  {journal} {\bibinfo  {journal}
			{J. Chem. Phys.}\ }\textbf {\bibinfo {volume} {80}},\ \bibinfo {pages} {2767}
		(\bibinfo {year} {1983})}\BibitemShut {NoStop}%
	\bibitem [{\citenamefont {Oliveira}\ and\ \citenamefont
		{Stilck}(2011)}]{doi:10.1063/1.3658045}%
	\BibitemOpen
	\bibfield  {author} {\bibinfo {author} {\bibfnamefont {T.~J.}\ \bibnamefont
			{Oliveira}}\ and\ \bibinfo {author} {\bibfnamefont {J.~F.}\ \bibnamefont
			{Stilck}},\ }\bibfield  {title} {\bibinfo {title} {Solution on the bethe
			lattice of a hard core athermal gas with two kinds of particles},\ }\href
	{https://doi.org/10.1063/1.3658045} {\bibfield  {journal} {\bibinfo
			{journal} {J. Chem. Phys.}\ }\textbf {\bibinfo {volume} {135}},\ \bibinfo
		{pages} {184502} (\bibinfo {year} {2011})}\BibitemShut {NoStop}%
	\bibitem [{\citenamefont {Oliveira}\ and\ \citenamefont
		{Stilck}(2015)}]{2015-os-pre-transfer}%
	\BibitemOpen
	\bibfield  {author} {\bibinfo {author} {\bibfnamefont {T.~J.}\ \bibnamefont
			{Oliveira}}\ and\ \bibinfo {author} {\bibfnamefont {J.~F.}\ \bibnamefont
			{Stilck}},\ }\bibfield  {title} {\bibinfo {title} {Transfer-matrix study of a
			hard-square lattice gas with two kinds of particles and density anomaly},\
	}\href {https://doi.org/10.1103/PhysRevE.92.032101} {\bibfield  {journal}
		{\bibinfo  {journal} {Phys. Rev. E}\ }\textbf {\bibinfo {volume} {92}},\
		\bibinfo {pages} {032101} (\bibinfo {year} {2015})}\BibitemShut {NoStop}%
	\bibitem [{\citenamefont {Liu}\ and\ \citenamefont
		{Evans}(2001)}]{liu2001phase}%
	\BibitemOpen
	\bibfield  {author} {\bibinfo {author} {\bibfnamefont {D.-J.}\ \bibnamefont
			{Liu}}\ and\ \bibinfo {author} {\bibfnamefont {J.~W.}\ \bibnamefont
			{Evans}},\ }\bibfield  {title} {\bibinfo {title} {Phase transitions in mixed
			adsorbed layers: Effect of repulsion between “hard squares” and “point
			particles”},\ }\href@noop {} {\bibfield  {journal} {\bibinfo  {journal} {J.
				Chem. Phys.}\ }\textbf {\bibinfo {volume} {114}},\ \bibinfo {pages} {10977}
		(\bibinfo {year} {2001})}\BibitemShut {NoStop}%
	\bibitem [{\citenamefont {Rodrigues}\ and\ \citenamefont
		{Oliveira}(2019{\natexlab{a}})}]{rodrigues2019three}%
	\BibitemOpen
	\bibfield  {author} {\bibinfo {author} {\bibfnamefont {N.~T.}\ \bibnamefont
			{Rodrigues}}\ and\ \bibinfo {author} {\bibfnamefont {T.~J.}\ \bibnamefont
			{Oliveira}},\ }\bibfield  {title} {\bibinfo {title} {Three stable phases and
			thermodynamic anomaly in a binary mixture of hard particles},\ }\href@noop {}
	{\bibfield  {journal} {\bibinfo  {journal} {J. Chem. Phys.}\ }\textbf
		{\bibinfo {volume} {151}},\ \bibinfo {pages} {024504} (\bibinfo {year}
		{2019}{\natexlab{a}})}\BibitemShut {NoStop}%
	\bibitem [{\citenamefont {Rodrigues}\ and\ \citenamefont
		{Oliveira}(2019{\natexlab{b}})}]{rodrigues2019thermodynamic}%
	\BibitemOpen
	\bibfield  {author} {\bibinfo {author} {\bibfnamefont {N.~T.}\ \bibnamefont
			{Rodrigues}}\ and\ \bibinfo {author} {\bibfnamefont {T.~J.}\ \bibnamefont
			{Oliveira}},\ }\bibfield  {title} {\bibinfo {title} {Thermodynamic behavior
			of binary mixtures of hard spheres: Semianalytical solutions on a husimi
			lattice built with cubes},\ }\href@noop {} {\bibfield  {journal} {\bibinfo
			{journal} {Phys. Rev. E}\ }\textbf {\bibinfo {volume} {100}},\ \bibinfo
		{pages} {032112} (\bibinfo {year} {2019}{\natexlab{b}})}\BibitemShut
	{NoStop}%
	\bibitem [{\citenamefont {Rodrigues}\ and\ \citenamefont
		{Oliveira}(2020)}]{rodrigues2020fluid}%
	\BibitemOpen
	\bibfield  {author} {\bibinfo {author} {\bibfnamefont {N.~T.}\ \bibnamefont
			{Rodrigues}}\ and\ \bibinfo {author} {\bibfnamefont {T.~J.}\ \bibnamefont
			{Oliveira}},\ }\bibfield  {title} {\bibinfo {title} {Fluid-fluid demixing and
			density anomaly in a ternary mixture of hard spheres},\ }\href@noop {}
	{\bibfield  {journal} {\bibinfo  {journal} {Phys. Rev. E}\ }\textbf {\bibinfo
			{volume} {101}},\ \bibinfo {pages} {062102} (\bibinfo {year}
		{2020})}\BibitemShut {NoStop}%
\end{thebibliography}
\end{document}